\documentclass[english]{article}
\usepackage[T1]{fontenc}
\usepackage[latin9]{luainputenc}
\usepackage{geometry}
\usepackage{amstext}
\usepackage{amssymb}
\usepackage{esint}
\usepackage{feyn}
\usepackage{tikz}
\usepackage{graphicx}
\usepackage[compat=1.1.0]{tikz-feynman}
\usepackage{array}
\usepackage{booktabs}

\usepackage[normalem]{ulem}
\usepackage{color}
\usepackage{listings}
\definecolor{darkgreen}{rgb}{0,0.35,0}
\definecolor{Rood}{rgb}{1, 0, 0}

\makeatletter
\usepackage{babel}
\usepackage{feyn}

\makeatother

\begin{document}

\title{\textbf{Study of the renormalization of BRST invariant local composite operators in the $U(1)$ Higgs model}}

\author{\textbf{M.~A.~L.~Capri$^1$}\thanks{caprimarcio@gmail.com},
\textbf{I.~F.~Justo$^2$}\thanks{igorfjusto@gmail.com}, 
\textbf{L.~F.~Palhares$^1$}\thanks{leticia.palhares@uerj.br},
\textbf{G.~Peruzzo$^1$}\thanks{gperuzzofisica@gmail.com},
\textbf{S.~P.~Sorella$^1$}\thanks{silvio.sorella@gmail.com},\\\\\
\textit{{\small $^1$UERJ -- Universidade do Estado do Rio de Janeiro,}}\\
\textit{{\small Instituto de F\'{\i}sica -- Departamento de F\'{\i}sica Te\'orica -- Rua S\~ao Francisco Xavier 524,}}\\
\textit{{\small 20550-013, Maracan\~a, Rio de Janeiro, Brasil}}\\
\textit{{\small $^2$UFF $-$  Universidade Federal Fluminense, }}\\
\textit{{\small Instituto de F\'{\i}sica, Campus da Praia Vermelha,}}\\
\textit{{\small Avenida General Milton Tavares de Souza s/n,}}\\
\textit{{\small 24210-346, Niter{\'o}i, RJ, Brasil}}\\\
}

\maketitle
\begin{abstract}
The renormalization properties of two local BRST invariant composite operators, $(O,V_\mu)$, corresponding respectively to the gauge invariant description of the Higgs particle and of the massive gauge vector boson, are scrutinized  in the $U(1)$ Higgs model by means of the algebraic renormalization setup. Their  renormalization $Z$'s factors are explicitly evaluated at one-loop order in the $\overline{\text{MS}}$ scheme by taking into due account the mixing with other gauge invariant operators. In particular, it turns out that the operator $V_\mu$ mixes with the gauge invariant quantity $\partial_\nu F_{\mu\nu}$, which has the same quantum numbers, giving rise to a $2 \times 2$ mixing matrix.  Moreover, two additional powerful Ward identities exist which enable us to determine the whole set of  $Z$'s factors entering the $2 \times 2$ mixing matrix as well as the $Z$ factor of the operator $O$ in a purely algebraic way. An  explicit check of these Ward identities is provided. The final setup obtained allows for computing perturbatively the full renormalized result for any $n$-point correlation function of the scalar and vector composite operators.
\end{abstract}

\section{Introduction}

In two previous works \cite{Dudal:2019aew,Dudal:2019pyg}, the elementary excitations of the $U(1)$ Higgs model, namely, the Higgs particle and the vector massive gauge boson, have been investigated within a fully gauge invariant setup, relying on the introduction of two local BRST invariant operators \cite{hooft2012nonperturbative,Frohlich:1980gj,Frohlich:1981yi} $(O,V_\mu)\footnote{See \cite{Maas:2019nso,Maas:2017xzh} and refs. therein for a recent account on the subject.} $: 
\begin{eqnarray} 
O\left(x\right) & = & \frac{1}{2}\left(h^{2}+2vh+\rho^{2}\right) \,, \nonumber \\
V_{\mu}\left(x\right) & = & \frac{1}{2}\left(-\rho\partial_{\mu}h+h\partial_{\mu}\rho+v\partial_{\mu}\rho+eA_{\mu}\left(v^{2}+h^{2}+2vh+\rho^{2}\right)\right) \,, \label{ovop}
\end{eqnarray} 
where $(h,\rho)$ stand for the Higgs and Goldstone fields, the parameter $v$ is the minimum of the classical Higgs potential, while $A_\mu$ is the gauge field and $e$ is the gauge coupling.\\\\More precisely, the explicit one-loop computation of the two-point correlation functions
\begin{equation} 
\langle O(p) O(-p) \rangle \;, \qquad \langle V_\mu(p) V_\nu(-p) \rangle \;, \label{corrf}
\end{equation} 
worked out in \cite{Dudal:2019pyg} in the 't Hooft $R_\xi$ gauge has revealed that, besides being independent from the gauge parameter $\xi$, the pole masses of $\langle O(p) O(-p) \rangle$ and $\langle V_\mu(p) V_\nu(-p) \rangle^{T} $ coincide, respectively, with the pole masses of the corresponding elementary correlation functions $\langle h(p) h(-p) \rangle$ and $\langle A_\mu(p) A_\nu(-p) \rangle^{T} $, where $\langle \; \; \rangle^{T}$ denotes the transverse components\footnote{The correlation functions $\langle A_\mu(p) A_\nu(-p) \rangle $ and $\langle V_\mu(p) V_\nu(-p) \rangle $ can be decomposed into transverse and longitudinal components as usual:
\begin{equation} 
\langle A_\mu(p) A_\nu(-p) \rangle = P_{\mu\nu} D(p^2) + L_{\mu\nu} L(p^2) \;, \qquad
\langle A_\mu(p) A_\nu(-p) \rangle^{T} = P_{\mu\nu} D(p^2) \;, \label{dec}  
\end{equation} 
where $P_{\mu\nu}=(\delta_{\mu\nu} - \frac{p_\mu p_\nu}{p^2}) $ and $L_{\mu\nu}=  \frac{p_\mu p_\nu}{p^2}$ are the transverse and longitudinal projectors.} of $\langle V_{\mu}(p) V_{\nu}(-p) \rangle $ and $\langle A_{\mu}(p) A_{\nu}(-p) \rangle $. Moreover, both tree-level and one-loop expressions for the longitudinal part of 
$\langle V_\mu(p) V_\nu(-p) \rangle $ remain independent from the momentum $p^2$ \cite{Dudal:2019pyg}, so that they are not associated to any physical mode, a feature which is expected to hold to higher orders.
\\\\Although the independence from the gauge parameter $\xi$ of the pole masses of $\langle h(p) h(-p) \rangle$ and $\langle A_\mu(p) A_\nu(-p) \rangle^{T} $ is ensured by the Nielsen identities \cite{Nielsen:1975fs,Piguet:1984js,Kraus:1995jk,Haussling:1996rq,Gambino:1998ec,Gambino:1999ai}, the corresponding residui are not protected in the same way. In particular, unlike\footnote{We remind here that the two-point correlation function $\langle A_\mu(p) A_\nu(-p) \rangle^{T}$ is  gauge invariant due to the gauge invariance  of  the transverse component of the Abelian gauge field $A^T_\mu = ( \delta_{\mu\nu} - \frac{\partial_\mu \partial_\nu}{\partial^2})A_\nu$.} what happens for $\langle A_\mu(p) A_\nu(-p) \rangle^{T} $, the residue of $\langle h(p) h(-p) \rangle$ turns out to be $\xi$-dependent \cite{Dudal:2019aew,Dudal:2019pyg}, a feature which originates from the lack of gauge invariance of the elementary Higgs field $h$.\\\\As a consequence, one finds that the spectral density of the K{\"a}ll{\'e}n-Lehmann (KL) representation of $\langle h(p) h(-p) \rangle$ depends on the gauge parameter $\xi$ as well, jeopardizing a direct and gauge invariant description of the Higgs particle in terms of the non-gauge invariant field $h$. From that perspective, the employment of the manifest gauge invariant operators $(O,V_\mu)$ enables us to overcome all the above mentioned difficulties. In fact, the correlator  $\langle O(p) O(-p) \rangle$ enjoys a KL representation with a well defined positive and $\xi$-independent spectral density, a property which holds also for  $\langle V_\mu(p) V_\nu(-p) \rangle^{T} $ \cite{Dudal:2019pyg}. As such, the correlation functions $\langle O(p) O(-p) \rangle$ and $\langle V_\mu(p) V_\nu(-p) \rangle^{T} $ provide a consistent gauge invariant description of the elementary excitations of the $U(1)$ Higgs model. It is worth remarking that the whole framework generalizes to the non-Abelian Yang-Mills-Higgs models \cite{hooft2012nonperturbative,Frohlich:1980gj,Frohlich:1981yi} as, for example, the $SU(2)$ Yang-Mills theory with a single Higgs field in the fundamental representation \cite{prep}. As one can easily figure out, in the non-Abelian case, besides the gauge dependence of $\langle h(p) h(-p) \rangle$, also the residue of the correlator $\langle A^a_\mu(p) A^b_\nu(-p) \rangle^{T} $ will exhibit a manifest $\xi$-dependence, due to the fact that in the non-Abelian case the transverse component of $A^a_\mu$, {\it i.e.} $A^{aT}_\mu = ( \delta_{\mu\nu} - \frac{\partial_\mu \partial_\nu}{\partial^2})A^a_\nu$, is no more gauge invariant. Therefore, in the non-Abelian case, the use of a fully gauge invariant setup relying on the non-Abelian generalization of $(O,V_\mu)$ \cite{hooft2012nonperturbative,Frohlich:1980gj,Frohlich:1981yi} turns out to be very welcome. 
\\\\The goal of this work is that of filling a gap not yet addressed in the previous analyses 
\cite{Dudal:2019aew,Dudal:2019pyg}, namely: the renormalization of the composite operators $(O,V_\mu)$ and of their correlation functions. To some extent, the present study completes the investigation of the operators  
$(O,V_\mu)$ in the $U(1)$ Higgs model, paving thus the path in order to face the more complex case of $SU(2)$, where non-perturbative issues related to the behaviour of the theory in the infrared region could be investigated within an exact BRST invariant framework.\\\\As is known from field theory textbooks, see \cite{Itzykson:1980rh}, the renormalization factor $Z_M$ of a given local operator $M(x)$, introduced into the starting action by means of an external source $J_M(x)$, {\it i.e.} $\int d^4x J_M(x) M(x)$,  can be extracted through the evaluation of the connected Green function\footnote{We remind that in the case in which a composite operator $M(x)$ is present in the theory, the generating functional ${\cal Z}^c(J_\phi,J_M)$ of the connected Green functions is defined through the Legendre transformation 
\begin{equation} 
{\cal Z}^c(J_\phi,J_M) = \Gamma(\phi,J_M) + \int d^4x J_\phi(x) \phi(x) \;, \label{zgg}
\end{equation} 
where $\Gamma(\phi,J_M)$ is the generator of the $1PI$ Green functions obtained by including the operator $M(x)$ in the starting action through the term $\int d^4 x J_M(x) M(x)$, where $J_M(x)$ is the external source needed to define the composite operator $M(x)$. Notice that, in eq.(\ref{zgg}), the Legendre transformation is taken only with respect to the variables $\{\phi\}$ and their corresponding sources $\{ J_\phi \}$, namely $J_\phi = - \frac{\delta \Gamma}{\delta \phi}$ and $\phi = \frac{\delta {\cal Z}^c}{\delta J_\phi}$.} 
\begin{equation}
\langle \phi(x_1)....\phi(x_n) M(y)\rangle = {\frac{\delta^{n+1} {\cal Z}^c(J_\phi,J_M) }{\delta J_\phi(x_1)...\delta J_\phi(x_n) \delta J_M(y)} }  \Biggl|_{(J_\phi,J_M)=0} \;, \label{an} 
\end{equation}
containing a suitable set of elementary fields $\{\phi(x)\}$ having a non-vanishing overlap with the insertion of the local composite operator $M(x)$ under investigation.\\\\Looking at the expressions (\ref{ovop}) of the two local composite operators $(O,V_\mu)$, it is apparent to realize that, due to the presence of terms linear in the fields $h$ and $A_\mu$, the simplest connected Green functions fulfilling the above-mentioned criterion are the two-point correlators: 
\begin{equation}
\langle h(x) O(y)\rangle = \frac{\delta^{2} {\cal Z}^c }{\delta J_h(x) \delta J(y) }  \Biggl|_{{\rm sources}=0}  \;, \label{1cc} 
\end{equation}
and 
\begin{equation}
\langle A_\mu(x) O(y)\rangle = \frac{\delta^{2} {\cal Z}^c }{\delta J^A_\mu(x) \delta \Omega_\nu(y) }  \Biggl|_{{\rm sources}=0}  \;, \label{2cc} 
\end{equation}
where $(J_h(x),J^A_\mu(x) )$ and $(J(x),\Omega_\mu(x) )$ are respectively the sources corresponding to the fields $(h(x),A_\mu(x))$ and to the composite operators $(O(x),V_\mu(x))$.
\\\\In the following sections, in order to extract the renormalization factors\footnote{We recall here that the renormalization $Z$'s factors  of local composite operators belonging to the cohomology of the BRST operator are independent from the gauge parameters entering the gauge fixing condition \cite{KlubergStern:1974rs,Joglekar:1975nu}, see also \cite{Dudal:2008tg,Piguet:1995er} and refs. therein.} of $O(x)$ and $V_\mu(x)$, we shall compute the correlators of eqs.(\ref{1cc}) and (\ref{2cc}) at one-loop order in the $\overline{\text{MS}}$ renormalization scheme. Moreover, resorting to the BRST invariant nature of $(O(x),V_\mu(x))$, we shall make use of the Landau gauge \cite{Clark:1974eq}, $\partial_\mu A_\mu=0$, which, in the present case, displays several practical advantages when compared to the $R_\xi$ gauge. In fact, besides the exact BRST invariance, the Landau gauge exhibits a manifest global $U(1)$ symmetry which implies a very useful relation among the renormalization factors of $(h(x),v, \rho(x))$, namely
\begin{equation}
    Z_h = Z_\rho = Z_v \;. \label{relz} 
\end{equation}
Such a relation is lost in the $R_\xi$ gauge, see \cite{Kraus:1995jk,Haussling:1996rq}. In addition, thanks to the transversality of the gauge condition $\partial_\mu A_\mu=0$, the correlator $\langle A_\mu(x) V_\nu(y)\rangle$ will be automatically projected into its transverse component.
\\\\The paper is organized as follows: in Section 2 we briefly review some basic features of the $U(1)$ Higgs model quantized in the Landau gauge. In Section 3 we shall introduce  the operators $(O(x),V_\mu(x))$. We shall first analyse them from the point of view of the cohomology of the BRST operator \cite{Piguet:1995er} in order to detect the existence of possible mixings with other operators. In particular, we shall see that $V_\mu(x)$ mixes with the gauge invariant operator $\partial_\nu F_{\mu\nu} =(\partial_\mu (\partial A) - \partial^2 A_\mu) $, while $O(x)$ requires the introduction of the constant quantity $v^2$ which is easily handled. We shall proceed then by introducing the starting BRST invariant classical action $\Sigma$ containing all needed operators and corresponding sources. The next step will be that of presenting the  Ward identities obeyed by $\Sigma$. In Section 4, following the algebraic renormalization setup \cite{Piguet:1995er}, we shall make use of the Ward identities to characterize the most general local invariant counterterm from which both bare action and renormalization $Z$'s factors will be introduced. Section 5 is devoted to the one-loop evaluation of the $Z$'s factors in the $\overline{\text{MS}}$ renormalization scheme, including those of the composite operators. The knowledge of the $Z$'s factors will allow us to provide some explicit checks of the consequences implied by the Ward identities of the $U(1)$ Higgs model in the Landau gauge. In particular, a whole subsection will be devoted to the existence of two special Ward identities, one local and one integrated, which enable us to characterize the $2 \times 2$ mixing matrix between the operators $(V_\mu,\partial_\nu F_{\mu\nu})$ as well as the $Z$ factor of the operator $O$ in a purely algebraic way. In Section 6 we present our conclusion and perspectives. The final Appendices contain  the calculations of the Feynman diagrams contributing to the correlation functions of eqs.(\ref{1cc}) and (\ref{2cc}). Finally, let us underline that a whole subsection of the Appendix \ref{B} has been devoted to the evaluation of the one-loop two point correlation function of the Goldstone field, $\langle \rho(p) \rho(-p)\rangle$, showing that it remains massless, as required by the global $U(1)$ Ward identity of the Landau gauge.

\section{Brief summary of the $U(1)$ Higgs model in the Landau gauge}

The  Abelian $U(1)$ Higgs model \cite{Higgs:1964pj,Higgs:1964ia,Englert:1964et,Guralnik:1964eu} is characterized by the following
action 

\begin{eqnarray}
S_{\rm Higgs} & = & \int d^{4}x\left[\frac{1}{4}F_{\mu\nu}F_{\mu\nu}+\left(D_{\mu}\varphi\right)^{\ast}\left(D_{\mu}\varphi\right)+\frac{1}{2}\lambda\left(\left|\varphi\right|^{2}-\frac{v^{2}}{2}\right)^{2}\right] \;, \label{ha}
\end{eqnarray}
where 

\begin{eqnarray}
F_{\mu\nu} & = & \partial_{\mu}A_{\nu}-\partial_{\nu}A_{\mu} \;, \nonumber \\
D_{\mu}\varphi & = & (\partial_{\mu}+ieA_{\mu})\varphi \;, \label{def}
\end{eqnarray}
with $\varphi$ being a complex scalar field, $e$  the electric charge and  $\lambda$ the quartic self-coupling.\\\\Expanding the complex field $\varphi$ around the minimum of the classical potential in eq.(\ref{ha}), {\it i.e.}
\begin{eqnarray*}
\varphi & = & \frac{1}{\sqrt{2}}\left(v+h+i\rho\right)\;, \label{exp}
\end{eqnarray*}
where $h$ and $\rho$ are the Higgs  and the Goldstone fields, expression (\ref{ha}) becomes 
\begin{eqnarray}
S_{\rm Higgs} & = & \int d^{4}x\left[\frac{1}{4}F_{\mu\nu}F_{\mu\nu}+\frac{1}{2}\partial_{\mu}h\partial_{\mu}h+\frac{1}{2}\partial_{\mu}\rho\partial_{\mu}\rho\right.\nonumber 
  +\frac{1}{2}e^{2}v^{2}A_{\mu}A_{\mu}+evA_{\mu}\partial_{\mu}\rho+\frac{1}{2}\lambda v^{2}h^{2}\nonumber \\
 &  & \;\;\;\;\;-eA_{\mu}\rho\partial_{\mu}h+eA_{\mu}h\partial_{\mu}\rho+e^{2}vhA_{\mu}A_{\mu}\nonumber  +\frac{1}{2}e^{2}\rho^{2}A_{\mu}A_{\mu}+\frac{1}{2}e^{2}h^{2}A_{\mu}A_{\mu}\nonumber \\
 &  & \;\;\;\;\;\left.+\frac{1}{8}\lambda h^{4}+\frac{1}{8}\lambda\rho^{4}+\frac{1}{2}\lambda vh^{3}+\frac{1}{2}\lambda vh\rho^{2}+\frac{1}{4}\lambda h^{2}\rho^{2}\right] \;, \label{ah2} 
\end{eqnarray}
showing that both gauge and Higgs fields have acquired a mass, given respectively by 
\begin{equation} 
m^2 =e^2 v^2  \;, \qquad m^2_h = \lambda v^2 \;. \label{mah} 
\end{equation} 
The field $\rho$, called the {\it would be Goldstone boson}, remains massless. The action (\ref{ah2}) is left invariant by the local gauge transformations 
\begin{equation}
\delta_\alpha A_{\mu}  =  -\partial_{\mu}\alpha \;, \qquad 
\delta_\alpha h =  -e\alpha \rho \;, \qquad \delta_\alpha \rho  =  e\alpha \left(v+h\right) \;, \label{omg}
\end{equation}
with $\alpha(x)$ a local gauge parameter: 
\begin{equation}
\delta_\alpha S_{\rm Higgs} = 0 \;. \label{ahinv}
\end{equation} 
In order to quantize the model, we employ the Landau gauge \cite{Clark:1974eq}, $\partial_\mu A_\mu=0$. Following the BRST procedure \cite{Piguet:1995er}, for the Landau gauge-fixing term we have 
\begin{eqnarray}
S_{\rm gf}  =  \int d^{4}x\left(ib\partial_{\mu}A_{\mu}+\overline{c}\partial^{2}c\right) \;, \label{lgfx}
\end{eqnarray}
where $b$ stands for the Nakanishi-Lautrup field, while $c$ and $\overline{c}$ are the Faddeev-Popov ghosts. The local gauge invariance, eq.(\ref{ahinv}), is now replaced by the exact nilpotent BRST invariance, namely 
\begin{equation} 
s\left(S_{\rm Higgs}+S_{\rm gf}\right)=0 \;,  \label{brsti} 
\end{equation} 
where
\begin{eqnarray}
sA_{\mu} & = & -\partial_{\mu}c \;, \qquad sc  =  0 \;, \nonumber \\
sh & = & -ec\rho \;, \qquad 
s\rho  =  ec\left(v+h\right) \;, \nonumber \\
s\overline{c} & = & ib \;, \qquad sb  =  0\;, \nonumber \\ 
s^2 & = & 0 \;.  \label{ss} 
\end{eqnarray}
Besides the BRST invariance, the action $\left(S_{\rm Higgs}+S_{\rm gf}\right)$ enjoys the discrete charge conjugation symmetry 
\begin{eqnarray}
A_{\mu} & \rightarrow & -A_{\mu} \;, \qquad  h  \rightarrow  h \; \nonumber \\
\rho & \rightarrow & -\rho \;, \qquad b  \rightarrow  -b \;, \nonumber \\
\overline{c} & \rightarrow &  -\overline{c} \;, \qquad   c \rightarrow  -c \;, \label{cgj} 
\end{eqnarray}
as well as the  global invariance 
\begin{eqnarray}
\delta_{\omega}h & = & -e\omega\rho \;, \qquad  \delta_{\omega}\rho =e\omega\left(v+h\right) 
\;, \nonumber \\
\delta_\omega A_\mu & = & 0\;, \qquad \delta_\omega \overline{c}=0  \;, \qquad \delta_\omega c = 0\;, \qquad \delta_\omega b= 0  \;, \label{gbo} 
\end{eqnarray}
with 
\begin{equation} 
\delta_{\omega}  \left(S_{\rm Higgs}+S_{\rm gf}\right)  =  0 \;, \label{cnst}
\end{equation}
where $\omega$ is a constant parameter. As we shall see in the following, the global invariance, eq.(\ref{cnst}), can be converted into a Ward identity which will imply helpful relationships between the various terms of the most general local invariant counterterm  needed to renormalize the model. It is worth observing here that, unlike in the $R_\xi$ gauge \cite{Kraus:1995jk,Haussling:1996rq}, the Faddeev-Popov ghosts 
$(\overline{c}, c)$ are now non-interacting fields, being completely decoupled. This is another useful advantage of the Landau gauge. The same property holds for the $b$-field, which appears only at the quadratic level. \\\\Let us end this short summary by noticing that the composite operators $(O(x),V_{\mu}(x))$ are, respectively, even and odd under charge conjugation, {\it i.e.}
\begin{eqnarray}
O(x) \ & \rightarrow O(x) \;, 
\nonumber \\
V_\mu(x) & \rightarrow - V_\mu(x) \;,  \label{odev}
\end{eqnarray}
a feature which will be exploited in the next section.

\section{Introduction of the BRST invariant operators ($O\left(x\right)$,$V_{\mu}\left(x\right)$)}

\subsection{BRST cohomology} 

In order to implement the set of Ward identities needed for the renormalization analysis of the composite operators $(O(x),V_{\mu}(x))$, let us first look at them by means of the cohomology of the BRST operator \cite{Piguet:1995er}. This study will enable us to detect possible mixings with other operators \cite{KlubergStern:1974rs,Joglekar:1975nu}. In ref. \cite{Dudal:2008tg}, for example, the full mixing matrix for the composite operator $F^a_{\mu\nu}(x) F^a_{\mu\nu}(x)$ has been worked out in the case of Yang-Mills theories.
\\\\ Let us start  with the scalar operator $O(x)$, eq.(\ref{ovop}). It has dimension two, ghost number zero\footnote{The fields $(A_\mu, h, \rho, b)$ have ghost number zero, while $(\overline{c}, c)$ have, respectively,  ghost number $(-1,1)$.} and is even under charge conjugation, eq.(\ref{odev}). Therefore, we look at the most general scalar non-integrated quantity with dimension two and even under charge conjugation, $\Delta\left(x\right)$, such that
$s\Delta\left(x\right)=0$ and $\Delta \neq s{\hat \Delta}$ for some ${\hat \Delta}$ with ghost number -1. After a little algebra, it turns out that the most general expression for $\Delta$ is given by
\begin{eqnarray}
\Delta\left(x\right) & = & a_1 O\left(x\right)+ a_2 v^{2} \;, \label{q2s}
\end{eqnarray}
where $a_1$ and $a_2$ are arbitrary coefficients. We see therefore that the introduction of the even operator $O(x)$ requires that of the quantity $v^2$, a task easily done due to  its constant nature. \\\\Let us consider then the case of the odd, dimension three vector operator $V_{\mu}(x)$, eq.(\ref{ovop}). We look thus at the most general odd vector quantity, $\Delta_{\mu}\left(x\right)$, of dimension three and ghost number zero such that $s\Delta_{\mu}\left(x\right)=0$. In the vector case, it turns out that the most general expression for $\Delta_{\mu}\left(x\right)$ is provided by 
\begin{eqnarray}
\Delta_{\mu}\left(x\right) & = & c_{1}V_{\mu}\left(x\right)+c_{2}\partial_{\nu}F_{\nu\mu}+c_{3}\partial_{\mu}b \;, \label{vecase}
\end{eqnarray}
where $(c_1,c_2,c_3)$ are arbitrary coefficients.  We notice that, as already mentioned, the gauge invariant quantity $\partial_{\nu}F_{\nu\mu}$ shows up.  It is worth mentioning here that the presence of this  term  was already pointed out long ago by Clark in his work on the Abelian Higgs model in the Landau gauge \cite{Clark:1974eq}. \\\\ The term $\partial_{\mu}b$ corresponds to the exact trivial part of the cohomolgy of the operator $s$, as $\partial_{\mu}b = -i s(\partial_{\mu}\overline{c})$. Moreover, due to the rich set of Ward identities exhibited by the model in Landau gauge, the term $\partial_{\mu}b$ will not give rise to contributions at the quantum level, a feature also corroborated by the fact that $b$ is not an interacting field. Finally, the two operators $(V_{\mu},\partial_{\nu}F_{\nu\mu})$ will give rise to a $2\times 2$ mixing matrix, encoded in the renormalization of the corresponding external sources  needed to introduce them.
\\\\We have now all ingredients to construct the most general BRST-invariant starting action to face the goal of the present work. This will be the issue addressed in the next section.

\subsection{The complete starting classical action $\Sigma$ and its Ward identities}

Having identified all quantities needed for the renormalization of the composite operators $(O(x),V_{\mu}(x))$, we proceed by introducing  the following external sources term 
\begin{equation} 
S_{\rm ext}  =  \int d^{4}x \left(L(sh)+R(s\rho)  + JO+\eta v^{2}+\Omega_{\mu}V_{\mu}+\Upsilon_{\mu}\partial_{\nu}F_{\nu\mu}+\Theta_{\mu}\partial_{\mu}b \right) \;. \label{ext}
\end{equation} 
The sources $(L,R)$ are needed in order to define the non-linear BRST variations of the fields $(h, \rho)$ \cite{Piguet:1995er}, eqs. (\ref{ss}). The remaining sources $(J,\eta,\Omega_{\mu},\Upsilon_{\mu},\Theta_{\mu})$ are needed to introduce all quantities which appear in the previous BRST cohomology analysis of $(O(x),V_{\mu}(x))$.  All external sources are BRST invariant, {\it i.e.} 
\begin{equation} 
sL= sR = sJ=s\eta=s\Omega_{\mu}=s\Upsilon_{\mu}=s\Theta_{\mu}=0 \;. \label{sext} 
\end{equation} 
For the complete form of the starting classical action $\Sigma$ we write thus 

\begin{eqnarray}
\Sigma & = & S_{\rm Higgs}+S_{\rm gf}+S_{\rm ext} \;, \label{cact} 
\end{eqnarray}
with 
\begin{equation} 
s \Sigma = 0 \;. \label{ssig}
\end{equation}
The fields $(A_\mu,h, \rho, b)$ have dimensions $(1,1,1,2)$ and ghost number zero. The Faddeev-Popov ghosts $(\overline{c},c) $ have dimensions $(2,0)$ and ghost number $(-1,1)$. The two external sources $(L, R)$ have dimension three and ghost number $-1$. Finally, the  sources $(J,\eta,\Omega_{\mu},\Upsilon_{\mu},\Theta_{\mu} )$ all have vanishing ghost number and dimensions $(2,2,1,1,1)$. 
\\\\It turns out that the complete classical action $\Sigma$ fulfills a huge number of Ward identities, which we enlist below:  
\begin{itemize}
\item  the Slavnov-Taylor identity expressing the BRST invariance of $\Sigma$ at the functional level 
\end{itemize}
\begin{eqnarray}
\mathcal{S}\left(\Sigma\right) & = & 0 \;, \label{slavnov}
\end{eqnarray}
where
\begin{eqnarray}
\mathcal{S}\left(\Sigma\right) & = & \int d^{4}x\left(-\partial_{\mu}c\frac{\delta\Sigma}{\delta A_{\mu}}+\frac{\delta\Sigma}{\delta L}\frac{\delta\Sigma}{\delta h}+\frac{\delta\Sigma}{\delta R}\frac{\delta\Sigma}{\delta\rho}+ib\frac{\delta\Sigma}{\delta\overline{c}}\right) \;. \label{st1}
\end{eqnarray}
\begin{itemize}
\item The $b$-Ward identity \cite{Piguet:1995er}
\end{itemize}
\begin{eqnarray}
\frac{\delta\Sigma}{\delta b} & = & i\partial_{\mu}A_{\mu}-\partial_{\mu}\Theta_{\mu}\;. \label{eq:bequation}
\end{eqnarray}
Notice that the right hand side of eq.(\ref{eq:bequation}), being linear in the quantum fields, is a linear breaking, not affected by quantum corrections \cite{Piguet:1995er}. This equation expresses in functional form the fact that the $b$ field is a non-interacting field.

\begin{itemize}
\item The antighost and ghost Ward identities 
\end{itemize}
\begin{eqnarray}
\frac{\delta\Sigma}{\delta\overline{c}} & = & \partial^{2}c\label{eq:barcequation} \;, 
\end{eqnarray}
and 
\begin{eqnarray}
\frac{\delta\Sigma}{\delta c} & = & -\partial^{2}\overline{c}-Re\left(v+h\right)+Le\rho\label{eq:cequation} \;. 
\end{eqnarray}
These two Ward identities express in functional form the decoupling of the Faddeev-Popov ghost fields in the Landau gauge. 
\begin{itemize}
\item The global invariance, eq.(\ref{gbo}), can be extended to the external sources in such a way that
\end{itemize}
\begin{eqnarray}
\delta_{\omega}h & = & -e\omega\rho \;, \qquad  \delta_{\omega}\rho =e\omega\left(v+h\right) 
\;, \nonumber \\
\delta_\omega A_\mu & = & 0\;, \qquad \delta_\omega \overline{c}=0  \;, \qquad \delta_\omega c = 0\;, \qquad \delta_\omega b= 0  \;, \nonumber \\ 
\delta_{\omega}L & = & -e\omega R\;, \qquad \delta_{\omega}R  =  e\omega L \;, \nonumber \\
\delta_\omega J &= & \delta_\omega \eta= \delta_\omega \Omega_{\mu}= \delta_\omega \Upsilon_{\mu}= \delta_\omega \Theta_{\mu}=0 \;, \label{gbos}
\end{eqnarray}
with 
\begin{equation} 
\delta_{\omega}  \Sigma  =  0 \;, \label{cnstinv}
\end{equation}
yielding the powerful Ward identity 
\begin{eqnarray}
\int d^{4}x\left[-\rho\frac{\delta\Sigma}{\delta h}+\left(v+h\right)\frac{\delta\Sigma}{\delta\rho}-R\frac{\delta\Sigma}{\delta L}+L\frac{\delta\Sigma}{\delta R}\right] & = & 0\label{eq:global} \;. \label{gboW}
\end{eqnarray}

\begin{itemize}
\item The charge conjugation  invariance
\end{itemize}
\begin{eqnarray}
A_{\mu} & \rightarrow & -A_{\mu} \;, \nonumber \\
h & \rightarrow & h\;, \nonumber \\
\rho & \rightarrow & -\rho \;, \nonumber \\
b & \rightarrow & -b \;, \nonumber \\
\overline{c} & \rightarrow & -\overline{c} \;, \nonumber \\
c & \rightarrow & -c \;, \nonumber \\
L & \rightarrow & L \;, \nonumber \\
R & \rightarrow & -R \;, \nonumber \\
J & \rightarrow & J \;, \nonumber \\
\Omega_{\mu} & \rightarrow & -\Omega_{\mu}\;, \nonumber \\
\Upsilon_{\mu} & \rightarrow & -\Upsilon_{\mu}\;, \nonumber \\
\Theta_{\mu} & \rightarrow & -\Theta_{\mu} \;. \label{wcc}
\end{eqnarray}
\begin{itemize}
\item The ghost number Ward identity  
\end{itemize}
\begin{eqnarray}
\mathcal{N}\left(\Sigma\right) & = & 0 \;, \label{eq:numberghost}
\end{eqnarray}
where

\begin{eqnarray}
\mathcal{N}\left(\Sigma\right) & = & \int d^{d}x\left(c\frac{\delta\Sigma}{\delta c}-\overline{c}\frac{\delta\Sigma}{\delta\overline{c}}-L\frac{\delta\Sigma}{\delta L}-R\frac{\delta\Sigma}{\delta R}\right) \;.
\end{eqnarray}
\begin{itemize}
\item The external sources Ward identities 
\end{itemize}
\begin{eqnarray}
\frac{\delta\Sigma}{\delta\eta} & = & v^{2} \;, \nonumber \\
\frac{\delta\Sigma}{\delta\Upsilon_{\mu}} & = & \partial_{\nu}F_{\nu\mu} \;, \nonumber \\
\frac{\delta\Sigma}{\delta\Theta_{\mu}} & = & \partial_{\mu}b\;.\label{seqs}
\end{eqnarray}
Notice that all terms in the right hand side of equations (\ref{seqs}) are linear breakings, which will not be affected by quantum corrections \cite{Piguet:1995er}. As a consequence, these equations  imply that the most general local invariant counterterm turns out to be independent from $(\eta,\Upsilon,\Theta )$.
\\\\ 
Let us end this section by observing that the power counting and the Ward identities could allow for terms purely quadratic in the sources $(J, \Omega_\mu)$ like, for example: 
\begin{equation}
\int d^4x\; J^2 \;, \qquad \int d^4x\; \Omega_\mu \partial^2 \Omega_\mu \;, \qquad {\rm etc,} 
\label{qd}
\end{equation} 
or even cubic and quartic ones. These terms would give rise to contact terms, see  \cite{Dudal:2019pyg}, in the correlation functions containing two (or more) insertions of the composite operators $(O(x),V_{\mu}(x))$ like: $\langle O(x) O(y)\rangle = \frac{\delta^2 {\cal Z}^c}{\delta J(x) \delta J(y)}\Big|_{\rm sources=0} $. These higher order terms are, however, not included in the complete starting classical action $\Sigma$, eq.(\ref{cact}),  since in a perturbative framework they, as well as any UV divergence related to them, arise from quantum corrections. Even though one cannot rule out nonperturbative tree-level contributions that could be associated e.g. with condensates involving the composite operators, we will not consider these phenomena and will attain ourselves to establishing a perturbative setup for treating the composite operators. \\\\All allowed nonlinear terms in the sources $(J, \Omega_\mu)$ will be identified and discussed in the next sections when writing down the most general counterterm. It should be noted thus that our final setup may be applied to thoroughly compute at the perturbative level any renormalized n-point correlation function of the composite operators.

\subsection{Extra Ward identities due to composite operators}
The set of Ward identities (\ref{slavnov})-(\ref{seqs}) can be written down  independently from the introduction of the composite operators $(O(x),V_{\mu}(x))$. It is worth thus opening a special subsection to mention specifically what happens when  $V_{\mu}$ and $O$ are coupled to the corresponding 
sources in the starting action. In the present work $V_{\mu}$ and
$O$ are the aim of our analysis, with our interest being the Green's functions involving 
 these composite operators. Let us therefore simply look at  $S_{\rm Higgs}$
and at the equations of motion of the fields $h$ and $A_{\mu}$ to try to figure out what kind of information they can provide. The action $S_{\rm Higgs}$ has
a quadratically broken global symmetry that holds $v$ and $h$ together, yielding 
\begin{eqnarray}
\int d^{4}x\left(\frac{\delta S_{\rm Higgs}}{\delta h}\right)-\frac{\partial S_{\rm Higgs}}{\partial v} & = & \int d^{4}x\lambda vO \;. \label{mar}
\end{eqnarray}
Notice that the right hand side of the previous equation contains precisely the composite operator $O(x)$. 
Expression (\ref{mar})  cannot be translated as it stands into a Ward identity at the quantum level unless the operator $O$ is introduced in the very beginning, as was already done in the case of $\Sigma$. If we replace $S_{\rm Higgs}$ by $\Sigma$, we get a true Ward identity:
\begin{eqnarray}
\int d^{4}x\left(\frac{\delta\Sigma}{\delta h}-\lambda v\frac{\delta\Sigma}{\delta J}\right)-\frac{\partial\Sigma}{\partial v} & = & \int d^{4}xv\left(J-2\eta\right).\label{hequation}
\end{eqnarray}
As already mentioned in the Introduction, besides the Ward identities (\ref{slavnov})-(\ref{seqs}) and (\ref{hequation}), the complete action $\Sigma$, eq.(\ref{cact}),  displays an additional local powerful identity which reads 
\begin{equation} 
\frac{\delta \Sigma}{\delta A_\mu} - 2 e \frac{\delta \Sigma}{\delta \Omega_\mu} - e \Omega_\mu \frac{\delta \Sigma}{\delta J} = -\partial_\nu F_{\nu\mu} - i \partial_\mu b + \frac{e v^2}{2} \Omega_\mu + \partial^2 \Upsilon_\mu - \partial_\mu \partial_\nu \Upsilon_{\nu} \;. \label{Widom}
\end{equation} 
The Ward identity (\ref{Widom}) relies on a rather nice feature of the vector operator $V_\mu(x)$. Let us look in fact again at the classical equations of motion which follow from the action $S_{\rm Higgs}$, eq.(\ref{ha}), namely 
\begin{equation} 
\frac{\delta S_{\rm Higgs}}{\delta A_\mu}   = - \left(\partial^2 \delta_{\mu\nu} - \partial_\mu \partial_\nu \right) A_\nu + 2e V_\mu 
\;, \label{eqmha}
\end{equation}
so that
\begin{equation}
\partial_\mu V_\mu = \frac{1}{2e} \partial_\mu  \frac{\delta S_{\rm Higgs}}{\delta A_\mu} \;, \label{curr}  
\end{equation}
which shows that, at the classical level, $V_\mu$ is a conserved current. The Ward identity (\ref{Widom}) expresses in a functional off-shell form this property of the operator $V_\mu$. Notice that, once more, the right hand side of (\ref{hequation}) and (\ref{Widom}) have a linear breaking, {\it i.e.} they are linear in the quantum fields, being unaffected by quantum corrections \cite{Piguet:1995er}.

\section{Algebraic characterization of the most general local invariant counterterm}

In order to characterize the most general local invariant counterterm, we follow the algebraic renormalization setup \cite{Piguet:1995er} and perturb the starting action $\Sigma$, {\it i.e.}
$\Sigma \rightarrow (\Sigma + \epsilon \Sigma^{\rm ct})$ with $\epsilon$ being an expansion parameter. In agreement with the power counting,  $\Sigma^{\rm ct}$ is 
an integrated local polynomial in the fields and linear in the external sources with dimension four, invariant under charge conjugation and having vanishing ghost number. Demanding then  that  the perturbed action, $(\Sigma + \epsilon \Sigma^{\rm ct})$, fulfills to first order in the expansion parameter $\epsilon$  the same Ward identities of the action $\Sigma$, namely eqs.(\ref{slavnov})-(\ref{seqs}), one gets the following conditions 
\begin{eqnarray}
\frac{\delta\Sigma^{\rm ct}}{\delta b} & = & \frac{\delta\Sigma^{\rm ct}}{\delta\overline{c}}=\frac{\delta\Sigma^{ct}}{\delta c}=0 \;, \label{cd1}
\end{eqnarray}
as well as 
\begin{eqnarray}
\frac{\delta\Sigma^{\rm ct}}{\delta\eta} = \frac{\delta\Sigma^{\rm ct}}{\delta\Theta_{\mu}}=\frac{\delta\Sigma^{ct}}{\delta\Upsilon_{\mu}}=0 \;.
\label{cd2} 
\end{eqnarray}
Since $\Sigma^{\rm ct}$ is independent from the antighost $\overline{c}$, it immediately follows that, due to the fact that the sources $(L,R)$ have ghost number $-1$, they cannot give rise to a dimension four quantity with vanishing ghost number, namely  
\begin{eqnarray}
\frac{\delta\Sigma^{\rm ct}}{\delta L} = \frac{\delta\Sigma^{\rm ct}}{\delta R}=0\;.\label{eq:sourcesequation}
\end{eqnarray}
Therefore 
\begin{equation} 
\Sigma^{\rm ct} = \Sigma^{ct}(A,h,\rho,v,J,\Omega) \;. \label{ct1} 
\end{equation} 
The result (\ref{eq:sourcesequation}) simplifies very much the Slavnov-Taylor
identity, which takes the simpler form
\begin{eqnarray}
s\Sigma^{\rm ct} = 0\;.
\end{eqnarray}
From equations (\ref{hequation}) and (\ref{Widom}) there are two additional conditions
\begin{eqnarray}
\frac{\delta\Sigma^{ct}}{\delta A_{\mu}}-2e\frac{\delta\Sigma^{ct}}{\delta\Omega_{\mu}}-e\Omega_{\mu}\frac{\delta\Sigma^{ct}}{\delta J} = 0
\label{aequationct}
\end{eqnarray}
and
\begin{eqnarray}
\int d^{4}x\left(\frac{\delta\Sigma^{ct}}{\delta h}-\lambda v\frac{\delta\Sigma^{ct}}{\delta J}\right)-\frac{\partial\Sigma^{ct}}{\partial v} = 0\,.
\label{hequationct}
\end{eqnarray}
After some algebraic calculations, it turns out that the most general form of $\Sigma^{ct}$ is given by 

\begin{eqnarray}
\Sigma^{ct} & = & \int d^{4}x\left\{ a_{0}\left(\frac{1}{4}F_{\mu\nu}F_{\mu\nu}-\frac{1}{2e}\Omega_{\mu}\partial_{\nu}F_{\nu\mu}-\frac{1}{8e^{2}}\Omega_{\mu}\partial^{2}\Omega_{\mu}+\frac{1}{8e^{2}}\Omega_{\mu}\partial_{\mu}\partial_{\nu}\Omega_{\nu}\right)\right. \nonumber\\
 &  & +a_{1}\left(\left(D_{\mu}\varphi\right)^{*}\left(D_{\mu}\varphi\right)+\Omega_{\mu}V_{\mu}+\frac{1}{8}v^{2}\Omega_{\mu}\Omega_{\mu}+\frac{1}{4}O\Omega_{\mu}\Omega_{\mu}\right) \nonumber \\
 &  & +a_{2}\left[\frac{\lambda}{2}\left(\varphi^{*}\varphi-\frac{v^{2}}{2}\right)^{2}+JO-\frac{1}{4}O\Omega_{\mu}\Omega_{\mu}+\frac{1}{32\lambda}\left(\Omega_{\mu}\Omega_{\mu}\Omega_{\nu}\Omega_{\nu}+16J^{2}-8J\Omega_{\mu}\Omega_{\mu}\right)\right] \nonumber \\
 &  & +\delta\sigma\left[\frac{v^{2}}{2}\left(h^{2}+2vh+\rho^{2}\right)+\frac{1}{\lambda}\left(Jv^{2}-\frac{1}{4}v^{2}\Omega_{\mu}\Omega_{\mu}-2JO+\frac{1}{2}O\Omega_{\mu}\Omega_{\mu}\right)\right. \nonumber \\
 &  & \left.\left.-\frac{1}{8\lambda^{2}}\left(\Omega_{\mu}\Omega_{\mu}\Omega_{\nu}\Omega_{\nu}+16J^{2}-8J\Omega_{\mu}\Omega_{\mu}\right)\right]\right\}
 \,,
 \label{eq:counterterm}
\end{eqnarray} 
where $(a_0, a_1, a_2,\delta \sigma)$ are free parameters. Expression (\ref{eq:counterterm}) displays a few features worth to be pointed out. The first one is the presence of the term $\Omega_{\mu}\partial_{\nu}F_{\mu\nu}$, with $\Omega_\mu$ being the source coupled to the vector operator $V_\mu$. As we shall see, the presence of this term gives rise to the mixing between the operators $V_\mu$ and $\partial_{\nu}F_{\mu\nu}$. The second feature concerns the BRST invariant counterterm $(\delta\sigma)\frac{v^{2}}{2} (h^2 +2vh +\rho^2)$. A quick inspection reveals that this term is not present in the starting classical action $\Sigma$. It has in fact been removed from $\Sigma$ by means of the expansion of the complex field $\varphi$, eq.(\ref{exp}),  around the minimum of the classical Higgs potential in eq.(\ref{ha}). However, the appearance of this term is a well known property of the Higgs model \cite{Becchi:1974md,Becchi:1974xu,Kraus:1995jk,Haussling:1996rq}, enabling us to cancel the tadpoles related to the Higgs field $h$, order per order in perturbation theory, {\it i.e.} it is determined by requiring the condition 
\begin{equation} 
\langle h\rangle = 0\;. \label{rad}
\end{equation}
Notice also the presence of the new source terms ($\Omega_{\mu}\partial^{2} \Omega_{\mu}$, $\Omega^4$, $J\Omega^2$,...). They are allowed not only by (\ref{cd1})-(\ref{eq:sourcesequation}), (\ref{aequationct}), (\ref{hequationct}) in $\Sigma^{ct}$, but also by (\ref{st1})-(\ref{eq:cequation}), (\ref{eq:global})-(\ref{eq:numberghost}), (\ref{seqs})-(\ref{Widom}) in the starting action $\Sigma$. However, as previously remarked, all of them, except possible condensate terms,  start from the order $\hbar$ onwards.\\\\Having obtained the most general form of the local invariant BRST counterterm, (\ref{eq:counterterm}), we proceed with the characterization of the bare action, namely 
\begin{eqnarray}
 \Sigma+\epsilon \Sigma^{\rm ct}= \Sigma_{\rm bare} + O(\epsilon^2) \;, \label{bare} 
\end{eqnarray}
 where 
\begin{eqnarray}
\Sigma_{\rm bare} & = & \Sigma\left(A_{0\mu},h_{0},\rho_{0},b_{0},c_{0},\overline{c}_{0},v_{0},e_{0},\lambda_{0},J_{0},\eta_{0},\Omega_{0\mu},\Upsilon_{0\mu},\Theta_{0\mu},L_{0},R_{0}\right) \nonumber \\
 &  & +\int d^{4}x\delta\sigma_{0}\frac{v_{0}^{2} }{2}\left(h_{0}^{2}+2v_{0}h_{0}+\rho_{0}^{2}\right) \nonumber \\
 &  & +\int d^{4}x\left\{ \left(Z_{A}-1\right)\left(-\frac{1}{8e_{0}^{2}}\Omega_{0\mu}\partial^{2}\Omega_{0\mu}+\frac{1}{8e_{0}^{2}}\Omega_{0\mu}\partial_{\mu}\partial_{\nu}\Omega_{0\nu}\right)\right. \nonumber \\
 &  & +\left(Z_{h}-1\right)\left(\frac{1}{8}v_{0}^{2}\Omega_{0\mu}\Omega_{0\mu}+\frac{1}{4}O_{0}\Omega_{0\mu}\Omega_{0\mu}\right)\nonumber \\
 &  & +\left(Z_{\lambda}+2Z_{h}-3\right)\left[-\frac{1}{4}O_{0}\Omega_{0\mu}\Omega_{0\mu}+\frac{1}{32\lambda_{0}}\left(\Omega_{0\mu}\Omega_{0\mu}\Omega_{0\nu}\Omega_{0\nu}+16J_{0}^{2}-8J_{0}\Omega_{0\mu}\Omega_{0\mu}\right)\right] \nonumber \\
 &  & +\delta\sigma_{0}\left[\frac{1}{\lambda_{0}}\left(-\frac{1}{4}v_{0}^{2}\Omega_{0\mu}\Omega_{0\mu}+\frac{1}{2}O_{0}\Omega_{0\mu}\Omega_{0\mu}\right)\right. \nonumber \\
 &  & \left.\left.-\frac{1}{8\lambda_{0}^{2}}\left(\Omega_{0\mu}\Omega_{0\mu}\Omega_{0\nu}\Omega_{0\nu}+16J_{0}^{2}-8J_{0}\Omega_{0\mu}\Omega_{0\mu}\right)\right]\right\} \label{bact}
\end{eqnarray}
with  
\begin{eqnarray}
A_{0\mu} & = & Z_{A}^{\frac{1}{2}}A_{\mu}\label{eq:renA} \;, \nonumber \\
h_{0} & = & Z_{h}^{\frac{1}{2}}h    \;, \nonumber \\
\rho_{0} & = & Z_{\rho}^{\frac{1}{2}}\rho \;, \nonumber \\
v_0 & = & Z_{v}^{\frac{1}{2}} v \;, \nonumber \\
b_{0} & = & Z_{b}^{\frac{1}{2}}b \;, \nonumber \\
c_{0} & = & Z_{c}^{\frac{1}{2}}c  \;, \nonumber \\
\overline{c}_{0} & = & Z_{\overline{c}}^{\frac{1}{2}}  \overline{c} \;, \nonumber \\
e_{0} & = & Z_{e}e \;, \nonumber \\
\lambda_{0} & = & Z_{\lambda}\lambda \;, \nonumber \\
L_{0} & = & Z_{L}L \;, \nonumber \\
R_{0} & = & Z_{R}R \;, \nonumber \\
\Theta_{\mu0} & = & Z_{\Theta}   \Theta \;, \label{z1}
\end{eqnarray}
and
\begin{eqnarray}
\left(\begin{array}{c}
\Omega_{0\mu}\\
\Upsilon_{0\mu}
\end{array}\right) & = & \left(\begin{array}{cc}
Z_{\Omega\Omega} & Z_{\Omega\Upsilon}\\
Z_{\Upsilon\Omega} & Z_{\Upsilon\Upsilon}
\end{array}\right)\left(\begin{array}{c}
\Omega_{\mu}\\
\Upsilon_{\mu}
\end{array}\right) \; \nonumber \\
\left(\begin{array}{c}
J_{0}\\
\eta_{0}
\end{array}\right) & = & \left(\begin{array}{cc}
Z_{JJ} & Z_{J\eta}\\
Z_{\eta J} & Z_{\eta\eta}
\end{array}\right)\left(\begin{array}{c}
J\\
\eta
\end{array}\right)\label{eq:renJ} \;. 
\end{eqnarray}
A simple inspection of equation (\ref{bare}) yields
\begin{eqnarray}
Z_{A}^{\frac{1}{2}} & = & Z_{e}^{-1}=1+\frac{1}{2}\epsilon  a_{0} \;, \nonumber \\
Z_{h}^{\frac{1}{2}} & = & Z_{\rho}^{\frac{1}{2}}=Z_{v}^{\frac{1}{2}}=1+\frac{1}{2}\epsilon a_{1} \;, \nonumber \\
Z_{\lambda} & = & 1+\epsilon \left(a_{2}-2a_{1}\right) \;, \nonumber \\
Z_{c}^{\frac{1}{2}} & = & Z_{\overline{c}}^{-\frac{1}{2}} \;, \nonumber \\
Z_{\Theta} & = & Z_{b}^{-\frac{1}{2}}=Z_{A}^{\frac{1}{2}}\;, \nonumber \\
Z_{L} & = & Z_{R}=Z_{e}^{-1}Z_{h}^{-\frac{1}{2}}Z_{c}^{-\frac{1}{2}} \;, \nonumber \\
Z_{\Omega\Omega} & = & 1\;, \nonumber \\
Z_{\Omega\Upsilon} & = & 0\;, \nonumber \\
Z_{\Upsilon\Omega} & = & -\frac{1}{2e}\epsilon  a_{0}= -\frac{1}{2e} \left(Z_{A}-1\right)\;, \nonumber \\
Z_{\Upsilon\Upsilon} & = & Z_{A}^{-\frac{1}{2}}=1-\frac{1}{2}\epsilon a_{0} \;, \nonumber \\
Z_{JJ} & = & 1+\epsilon \left(a_{2}-a_{1}-2\frac{\delta \sigma}{\lambda}\right) \;, \nonumber \\
Z_{J\eta} & = & 0 \;, \nonumber \\
Z_{\eta J} & = & \epsilon  \frac{\delta \sigma}{\lambda} \;, \nonumber \\
Z_{\eta\eta} & = & Z_{h}^{-1}=1-\epsilon  a_{1} \;, \label{zzz1}
\end{eqnarray}
and 
\begin{equation} 
(\delta \sigma)_0 = \epsilon (\delta \sigma) \;. \label{ds}
\end{equation} 
Thus, for the bare action, we get\footnote{Since we are not interested in the calculation of Green's functions with insertions of the BRST exact operators $(sh, s\rho)$, from now on, we shall set to zero  the corresponding external sources, {\it i.e.} 
$L=R=0$.} 

\begin{eqnarray}
\Sigma_{bare} & = & \int d^{4}x\left( \frac{1}{4}F_{0\mu\nu}F_{0\mu\nu}+\left(D_{0\mu}\varphi_{0}\right)^{\ast}\left(D_{0\mu}\varphi_{0}\right)+\frac{\lambda_{0}}{2}\left(\varphi_{0}^{\ast}\varphi_{0}-\frac{v_{0}^{2}}{2}\right)^{2} \right) \nonumber \\
 &  & + \int d^4x\; \left( \overline{c}_{0}\partial^{2}c_{0}+ib_{0}\partial_{\mu}A_{0\mu} +J_{0}O_{0}+\eta_{0}v_{0}^{2}
  +\Omega_{0\mu}V_{0\mu}+\Upsilon_{0\mu}\partial_{\nu}F_{0\nu\mu} \right) \nonumber \\
 &  & +  \int d^4x \left( \frac{(\delta \sigma)_0}{2} v^{2}_0\left(h^{2}_0+2v_0h_0+\rho^{2}_0\right) \right) \; \nonumber \\
 & &+\int d^{4}x\left\{ \left(Z_{A}-1\right)\left(-\frac{1}{8e_{0}^{2}}\Omega_{0\mu}\partial^{2}\Omega_{0\mu}+\frac{1}{8e_{0}^{2}}\Omega_{0\mu}\partial_{\mu}\partial_{\nu}\Omega_{0\nu}\right)\right. \nonumber \\
 &  & +\left(Z_{h}-1\right)\left(\frac{1}{8}v_{0}^{2}\Omega_{0\mu}\Omega_{0\mu}+\frac{1}{4}O_{0}\Omega_{0\mu}\Omega_{0\mu}\right)\nonumber \\
 &  & +\left(Z_{\lambda}+2Z_{h}-3\right)\left[-\frac{1}{4}O_{0}\Omega_{0\mu}\Omega_{0\mu}+\frac{1}{32\lambda_{0}}\left(\Omega_{0\mu}\Omega_{0\mu}\Omega_{0\nu}\Omega_{0\nu}+16J_{0}^{2}-8J_{0}\Omega_{0\mu}\Omega_{0\mu}\right)\right] \nonumber \\
 &  & +\delta\sigma_{0}\left[\frac{1}{\lambda_{0}}\left(-\frac{1}{4}v_{0}^{2}\Omega_{0\mu}\Omega_{0\mu}+\frac{1}{2}O_{0}\Omega_{0\mu}\Omega_{0\mu}\right)\right. \nonumber \\
 &  & \left.\left.-\frac{1}{8\lambda_{0}^{2}}\left(\Omega_{0\mu}\Omega_{0\mu}\Omega_{0\nu}\Omega_{0\nu}+16J_{0}^{2}-8J_{0}\Omega_{0\mu}\Omega_{0\mu}\right)\right]\right\} \label{bact}
 , \label{baction} 
\end{eqnarray}
with 
\begin{equation}
\varphi_0  =  \frac{Z_h^{1/2}}{\sqrt{2}}\left(v+h+i\rho\right)\;. \label{phi0}
\end{equation}
Equation (\ref{baction}) shows that, apart from the term $\epsilon \frac{\delta \sigma}{2} v^{2}_{0} (h^{2}_{0} + 2v_{0}h_{0} + \rho^{2}_{0}) $ 
and the purely source\footnote{It is worth observing also that the higher order terms in the external sources could be reabsorbed into the starting action $\Sigma$, eq.(\ref{cact}), by means of a non-linear redefinition of $(J,\eta)$, given by 
\begin{equation} 
J_0=J+\epsilon (z_{J}J+z_{\Omega} \Omega_{\mu}\Omega_{\mu}+z_{3}v^{2}) \;, \label{nrf1} 
\end{equation} 
and 
\begin{equation} 
\eta_{0}=\eta+\epsilon [z_{\eta}\eta+\tilde{z}_{J}J+\tilde{z}_{\Omega}\Omega_{\mu}\Omega_{\mu}+\frac{z_{4}}{v^{2}}(\Omega^{4}+16J^{2}-8J\Omega^{2})] \;, \label{nrf2}
\end{equation} 
with 
\begin{eqnarray} 
z_J & = &  a_{2}-a_{1}-2\frac{\delta\sigma}{\lambda}   \;, \qquad 
z_{\Omega}  =  -\frac{a_{2}}{4} \;, \nonumber \\
z_{3} & = & \frac{\delta\sigma}{2} \;, \qquad 
z_{\eta}  =  -a_{1} \;, \nonumber \\
\tilde{z}_{J} & = & \frac{\delta\sigma}{\lambda} \;, \qquad 
\tilde{z}_{\Omega}  =  -\frac{\delta\sigma}{4\lambda} \;, \qquad 
z_{4}=\frac{a_{2}}{32\lambda}-\frac{\delta\sigma}{8\lambda^{2}} \;.  \label{nlzf} 
\end{eqnarray} 
Notice that $(z_J,z_\eta)$ are the same expressions entering in the factors $Z_{JJ}$ and $Z_{\eta\eta}$ of 
eqs.(\ref{zzz1}).
} terms which, in a loop expansion, start at one-loop order onwards, the remaining terms of the general 
invariant counterterm can be reabsorbed  into the starting action (\ref{cact}) through a suitable redefinition of the fields, 
parameters and sources, establishing thus the already known renormalizability of the model, which is extended here with the 
introduction of the composite operators $(O(x), V_{\mu}(x))$. This final bare action represents a closed setup for 
perturbatively computing renormalized correlation functions involving any number of composite-operator insertions. \\\\Before 
starting with the one-loop evaluation of all $Z$'s factors and of $\delta\sigma$, let us point out a few features displayed by
eqs.(\ref{zzz1}):
\begin{itemize} 
\item as it is apparent from eqs.(\ref{zzz1}), the quantities $(v,h,\rho)$ have a common renormalization factor, {\it i.e.}
\begin{equation} 
Z_{h} =  Z_{\rho}=Z_{v} \;. \label{szf}
\end{equation}
This property follows from the rich set of Ward identities present in the Landau gauge, in particular from the existence of the global Ward identity (\ref{gboW}). The relation (\ref{szf}) turns out to be very helpful in the practical calculations of the $Z$'s factors of the composite operators $(O(x), V_\mu(x))$.
\item The renormalization factor of the electric charge $e$ is not independent from the renormalization factor of the gauge field $A_\mu$. This property, usually written as 
\begin{equation} 
 e_0 A_{0\mu} = e A_\mu \;, \qquad Z_e Z_A^{1/2} = 1 \;, \label{ea} 
\end{equation} 
is a well known feature of the Abelian U(1) models, being present also in spinor QED\footnote{This property is expressed, for example, in page 346, Eq.(7-73), of Ref. \cite{Itzykson:1980rh}.}. An explicit check of this property will be provided by the evaluation of the divergent one-loop contribution to the gauge boson mass $m$. 
\item 
Finally, let us rewrite the mixing matrices in a more explicit form, namely 
\begin{eqnarray}
\left(\begin{array}{c}
\Omega_{0\mu}\\
\Upsilon_{0\mu}
\end{array}\right) & = & \left(\begin{array}{cc}
1 & 0\\
-\frac{1}{2e}\epsilon a_0 & (1-\frac{\epsilon}{2}a_0) 
\end{array}\right)\left(\begin{array}{c}
\Omega_{\mu}\\
\Upsilon_{\mu}
\end{array}\right) \; \nonumber \\
\left(\begin{array}{c}
J_{0}\\
\eta_{0}
\end{array}\right) & = & \left(\begin{array}{cc}
(1+\epsilon(a_2-a_1-2\frac{\delta \sigma}{\lambda})) & 0\\
\epsilon \frac{\delta \sigma}{\lambda} & (1-\epsilon a_1)
\end{array}\right)\left(\begin{array}{c}
J\\
\eta
\end{array}\right)\label{mr} \;, 
\end{eqnarray}
from which one recognizes the general pattern given in \cite{KlubergStern:1974rs,Joglekar:1975nu}, see also \cite{Dudal:2008tg}. 
\end{itemize}

\noindent
We can now proceed with the one-loop evaluation of the $Z$'s factors.

\section{Explicit evaluation of the $Z$'s factors at one-loop order and check of
the Ward identities}
Having at our disposal the bare action, eq.(\ref{baction}), we can immediately obtain the one-loop action, including all needed counterterms, to face the evaluation of the  $Z$'s factors. Setting 
\begin{eqnarray}
Z_{A} & = & 1+\hbar Z_{A}^{\left(1\right)} \;, \nonumber \\
Z_{h} & = & 1+\hbar Z_{h}^{\left(1\right)} \;, \nonumber \\
Z_{\lambda} & = & 1+\hbar Z_{\lambda}^{\left(1\right)}\;,\nonumber  \\
Z_{JJ} & = & 1+ \hbar Z_{JJ}^{\left(1\right)}\;, \nonumber \\
Z_{\eta J} & = & \hbar Z_{\eta J}^{\left(1\right)} \;, \nonumber \\
Z_{J\eta} & = & 0\;, \nonumber \\
Z_{\eta\eta} & = & Z_{h}^{-1} \;, \nonumber \\
Z_{\Omega\Omega} & = & 1+\hbar Z_{\Omega\Omega}^{\left(1\right)}\;, \nonumber \\
Z_{\Upsilon\Omega} & = &\hbar  Z_{\Upsilon\Omega}^{\left(1\right)}\;, \nonumber \\
Z_{\Omega\Upsilon} & = & 0\;, \nonumber \\
Z_{\Upsilon\Upsilon} & = & Z_{A}^{-\frac{1}{2}}\;, \nonumber \\
(\delta\sigma)_{0} & = & \hbar (\delta\sigma)^{\left(1\right)} \;, \label{zexp} 
\end{eqnarray}
we get 
\begin{eqnarray}
\Sigma_{\rm bare} & = & \int d^{4}x \left\{ \frac{1}{4}\left(1+\hbar Z_{A}^{\left(1\right)}\right)F_{\mu\nu}F_{\mu\nu}+\left(1+ \hbar Z_{h}^{\left(1\right)}\right)\left(D_{\mu}\varphi\right)^{\ast}\left(D_{\mu}\varphi\right)\right.\nonumber \\
 &  & +\left(1+\hbar Z_{\lambda}^{\left(1\right)}\right)\left(1+2 \hbar Z_{h}^{\left(1\right)}\right)\frac{\lambda}{2}\left(\varphi^{\ast}\varphi-\frac{v^{2}}{2}\right)^{2}  \nonumber \\
 &  & +\overline{c}\partial^{2}c+ib\partial_{\mu}A_{\mu}
  +\hbar (\delta \sigma)^{\left(1\right)}\frac{1}{2}v^{2}\left(h^{2}+2vh+\rho^{2}\right)\nonumber \\
 &  & +\left(1+\hbar Z_{JJ}^{\left(1\right)}\right)\left(1+\hbar Z_{h}^{\left(1\right)}\right)JO+\hbar Z_{\eta J}^{\left(1\right)}Jv^{2}+\eta v^{2}  \nonumber \\
 &  & +\left(1+\hbar Z_{\Omega\Omega}^{\left(1\right)}\right)\left(1+\hbar Z_{h}^{\left(1\right)}\right)\Omega_{\mu}V_{\mu}+\hbar Z_{\Upsilon\Omega}^{\left(1\right)}\Omega_{\mu}\partial_{\nu}F_{\nu\mu}
 +\Upsilon_{\mu}\partial_{\nu}F_{\nu\mu} \Biggl\} \nonumber\\
 & &+\int d^{4}x\left\{\hbar Z_{A}^{\left(1\right)}\left(-\frac{1}{8e^{2}}\Omega_{\mu}\partial^{2}\Omega_{\mu}+\frac{1}{8e^{2}}\Omega_{\mu}\partial_{\mu}\partial_{\nu}\Omega_{\nu}\right)\right.\nonumber \\
 &  & +\hbar Z_{h}^{\left(1\right)}\left(\frac{1}{8}v^{2}\Omega_{\mu}\Omega_{\mu}+\frac{1}{4}O\Omega_{\mu}\Omega_{\mu}\right) \nonumber\\
 &  & +\hbar \left(Z_{\lambda}^{\left(1\right)}+2Z_{h}^{\left(1\right)}\right)\left[-\frac{1}{4}O\Omega_{\mu}\Omega_{\mu}+\frac{1}{32\lambda}\left(\Omega_{\mu}\Omega_{\mu}\Omega_{\nu}\Omega_{\nu}+16J^{2}-8J\Omega_{\mu}\Omega_{\mu}\right)\right] \nonumber \\
 &  & +\hbar \left(\delta\sigma\right)^{\left(1\right)}\left[\frac{1}{\lambda}\left(-\frac{1}{4}v^{2}\Omega_{\mu}\Omega_{\mu}+\frac{1}{2}O\Omega_{\mu}\Omega_{\mu}\right)\right. \nonumber \\
 &  & \left.\left.-\frac{1}{8\lambda^{2}}\left(\Omega_{\mu}\Omega_{\mu}\Omega_{\nu}\Omega_{\nu}+16J^{2}-8J\Omega_{\mu}\Omega_{\mu}\right)\right]\right\}
 +\;\; O(\hbar^2)\;.  \label{cct1}
\end{eqnarray}
Since
\begin{eqnarray}
\left(D_{\mu}\varphi\right)^{\ast}\left(D_{\mu}\varphi\right) & = & \frac{1}{2}\left[\left(\partial_{\mu}h\right)\left(\partial_{\mu}h\right)+\left(\partial_{\mu}\rho\right)\left(\partial_{\mu}\rho\right)+2evA_{\mu}\left(\partial_{\mu}\rho\right)+e^{2}v^{2}A_{\mu}A_{\mu}\right.\nonumber \\
 &  & \left.-2e\rho A_{\mu}\left(\partial_{\mu}h\right)+2ehA_{\mu}\left(\partial_{\mu}\rho\right)+e^2\rho^{2}A_{\mu}A_{\mu}+2e^{2}vhA_{\mu}A_{\mu}+e^{2}h^{2}A_{\mu}A_{\mu}\right] \;, \label{calc}
\end{eqnarray}
and
\begin{eqnarray}
\frac{\lambda}{2}\left(\varphi^{\ast}\varphi-\frac{v^{2}}{2}\right)^{2} & = & \frac{\lambda}{8}\left(h^{4}+4v^{2}h^{2}+\rho^{4}+4vh^{3}+2h^{2}\rho^{2}+4vh\rho^{2}\right) \;, \label{calc2} 
\end{eqnarray}
we can split the action into the sum of the quadratic piece with the interaction and one-loop counterterm, namely 
\begin{eqnarray*}
\Sigma_{\rm bare} & = & S^{\rm quad}+\hbar S_{\rm I} + O(\hbar^2) \;, \label{splitt}
\end{eqnarray*}
where
\begin{eqnarray}
S^{\rm quad} & = & \int d^{4}x\left\{ \frac{1}{2}A_{\mu}\left(-\delta_{\mu\nu}\partial^{2}+\partial_{\mu}\partial_{\nu}+m^{2}\delta_{\mu\nu}\right)A_{\nu}+ib\partial_{\mu}A_{\mu}\right.\nonumber \\
 &  & \left.+\frac{1}{2}h\left(-\partial^{2}+m_{h}^{2}\right)h+\frac{1}{2}\rho\left(-\partial^{2}\right)\rho+evA_{\mu}\left(\partial_{\mu}\rho\right)+\overline{c}\partial^{2}c\right\} \;, \label{qq} 
\end{eqnarray}
and 
\begin{eqnarray}
S_{\rm I} & = & \int d^{4}x\left\{ \frac{1}{2}Z_{A}^{\left(1\right)}A_{\mu}\left(-\delta_{\mu\nu}\partial^{2}+\partial_{\mu}\partial_{\nu}\right)A_{\nu}\right.\nonumber \\
 &  & +\frac{1}{2}Z_{h}^{\left(1\right)}\left[-2e\rho A_{\mu}\left(\partial_{\mu}h\right)+2ehA_{\mu}\left(\partial_{\mu}\rho\right)+e^2\rho^{2}A_{\mu}A_{\mu}+2e^{2}vhA_{\mu}A_{\mu}+e^{2}h^{2}A_{\mu}A_{\mu}\right]\nonumber \\
 &  & +\frac{1}{2}Z_{h}^{\left(1\right)}\left[-h\partial^{2}h-\rho\partial^{2}\rho+2evA_{\mu}\left(\partial_{\mu}\rho\right)+e^{2}v^{2}A_{\mu}A_{\mu}\right.\nonumber \\
 &  & \left.-2e\rho A_{\mu}\left(\partial_{\mu}h\right)+2ehA_{\mu}\left(\partial_{\mu}\rho\right)+e\rho^{2}A_{\mu}A_{\mu}+2e^{2}vhA_{\mu}A_{\mu}+e^{2}h^{2}A_{\mu}A_{\mu}\right]\nonumber \\
 &  & +\left(Z_{\lambda}^{\left(1\right)}+2 Z_{h}^{\left(1\right)}\right)\frac{\lambda}{2}v^{2}h^{2}\nonumber \\
 &  & +\left(1+\left(Z_{\lambda}^{\left(1\right)}+2 Z_{h}^{\left(1\right)}\right)\right)\frac{\lambda}{8}\left[h^{4}+\rho^{4}+4vh^{3}+2h^{2}\rho^{2}+4vh\rho^{2}\right]\nonumber \\
 &  & +(\delta\sigma)^{\left(1\right)}\frac{1}{2}v^{2}\left(h^{2}+2vh+\rho^{2}\right)\nonumber \\
 &  & +\frac{1}{2}\left(1+\left(Z_{JJ}^{\left(1\right)}+Z_{h}^{\left(1\right)}\right)\right)J\left(h^{2}+2vh+\rho^{2}\right)+\hbar Z_{\eta J}^{\left(1\right)}Jv^{2}+\eta v^{2}\nonumber \\
 &  & +\frac{1}{2}\left(1+\left(Z_{\Omega\Omega}^{\left(1\right)}+Z_{h}^{\left(1\right)}\right)\right)\Omega_{\mu}\left[-\rho\partial_{\mu}h+h\partial_{\mu}\rho+v\partial_{\mu}\rho+eA_{\mu}\left(v^{2}+h^{2}+2vh+\rho^{2}\right)\right]\nonumber \\
 &  & +Z_{\Upsilon\Omega}^{\left(1\right)}\Omega_{\mu}\partial_{\nu}F_{\nu\mu}+\Upsilon_{\mu}\partial_{\nu}F_{\nu\mu} \Biggl\} \;\nonumber\\
 & &+\int d^{4}x\left\{ Z_{A}^{\left(1\right)}\left(-\frac{1}{8e^{2}}\Omega_{\mu}\partial^{2}\Omega_{\mu}+\frac{1}{8e^{2}}\Omega_{\mu}\partial_{\mu}\partial_{\nu}\Omega_{\nu}\right)\right.\nonumber \\
 &  & + Z_{h}^{\left(1\right)}\left(\frac{1}{8}v^{2}\Omega_{\mu}\Omega_{\mu}+\frac{1}{4}O\Omega_{\mu}\Omega_{\mu}\right) \nonumber\\
 &  & +\left(Z_{\lambda}^{\left(1\right)}+2Z_{h}^{\left(1\right)}\right)\left[-\frac{1}{4}O\Omega_{\mu}\Omega_{\mu}+\frac{1}{32\lambda}\left(\Omega_{\mu}\Omega_{\mu}\Omega_{\nu}\Omega_{\nu}+16J^{2}-8J\Omega_{\mu}\Omega_{\mu}\right)\right] \nonumber \\
 &  & +\left(\delta\sigma\right)^{\left(1\right)}\left[\frac{1}{\lambda}\left(-\frac{1}{4}v^{2}\Omega_{\mu}\Omega_{\mu}+\frac{1}{2}O\Omega_{\mu}\Omega_{\mu}\right)\right. \nonumber \\
 &  & \left.\left.-\frac{1}{8\lambda^{2}}\left(\Omega_{\mu}\Omega_{\mu}\Omega_{\nu}\Omega_{\nu}+16J^{2}-8J\Omega_{\mu}\Omega_{\mu}\right)\right]\right\}. 
  \label{eq:sinter}
\end{eqnarray}
From expressions (\ref{qq}) and (\ref{eq:sinter}) we can derive the tree level field propagators and the one-loop Feynman rules obtained by keeping the sources $(J, \eta,\Omega_\mu, \Upsilon_\mu)$ as external fields.\\\\
For the benefit of the reader, all propagators and Feynman rules can be found in  Appendix \ref{A}. Appendix \ref{B} collects all details of the evaluation of the Green functions of the elementary fields, {\it i.e.} $\langle h \rangle$, $\langle h(x) h(y)\rangle$, $\langle A_\mu(x) A_\nu(y)\rangle$. A complete subsection has also been devoted to the one-loop calculation of the Goldstone two-point Green function $\langle \rho(x) \rho(y) \rangle$, showing that  the Goldstone mode remains massless in the Landau gauge, providing a very useful check of the whole setup. Appendix \ref{C} is fully devoted to the details of the evaluation of the Green functions of the composite operators  $\left\langle h\left(x\right)O\left(y\right)\right\rangle $
and $\left\langle A_{\mu}\left(x\right)V_{\nu}\left(y\right)\right\rangle $. \\\\Let us start with the one-loop vanishing tadpoles condition 
\begin{eqnarray}
\left\langle h\left(x\right)\right\rangle_{\rm 1-loop}  = 0 \;. \label{tdp1}
\end{eqnarray}
From eqs.(\ref{tdpcms})-(\ref{divtad}) of Appendix \ref{B}, one obtains
\begin{eqnarray}
(\delta\sigma)^{\left(1\right)}  =  \frac{1}{v^{2}}\left(-e^{2}\left(d-1\right)\chi\left(m^{2}\right)-\frac{3}{2}\lambda\chi\left(m_{h}^{2}
\right)\right) \;, \label{tdp2}
\end{eqnarray}
and, for the one-loop divergent part of $(\delta\sigma)^{\left(1\right)}$ in the $\overline{\text{MS}}$ scheme (with $d=4-\varepsilon$): 
\begin{eqnarray}
 (\delta\sigma)^{\left(1\right)}_{\rm div} =  \frac{1}{\left(4\pi\right)^{2}}\frac{1}{v^{2}}\left(3e^{2}m^{2}+\frac{3}{2}\lambda m_{h}^{2}\right)
\left(\frac{2}{\varepsilon}-\gamma+\ln\left(4\pi\right)\right) \;. \label{tdp3}
\end{eqnarray}
The renormalization factors  $(Z_{A}^{\left(1\right)},Z_{h}^{\left(1\right)},Z_{\lambda}^{\left(1\right)})$ 
can be obtained evaluating the one-loop connected two-point functions $\left\langle A_{\mu}\left(x\right)A_{\nu}\left(y\right)\right\rangle $,  $\left\langle h\left(x\right)h\left(y\right)\right\rangle $ and $\left\langle \rho\left(x\right)\rho\left(y\right)\right\rangle $, all presented in Appendix \ref{B}. It turns out that\footnote{The factors $(Z_{A}^{(1)},Z_{\lambda}^{(1})$ give rise to the standard  expressions for the one-loop $\beta$ functions \cite{Kraus:1995jk}, namely
\begin{equation}
\beta^{(1)}_{e} = \frac{e^3}{48\pi^2}\;, \qquad \beta^{(1)}_\lambda= \frac{1}{8\pi^2}(5 \lambda^2 -6 e^2\lambda + 6 e^4) \;. \label{beta-func} 
\end{equation}
}
\begin{eqnarray}
Z_{A}^{\left(1\right)} & = & -\frac{e^{2}}{48\pi^{2}}\left(\frac{2}{\varepsilon}-\gamma+\ln\left(4\pi\right)\right) \;, \label{zaat}
\end{eqnarray}
\begin{eqnarray}
Z_{h}^{\left(1\right)} & = & \frac{3e^{2}}{16\pi^{2}}\left(\frac{2}{\varepsilon}-\gamma+\ln\left(4\pi\right)\right) \;, \label{zhht}
\end{eqnarray}
\begin{eqnarray}
Z_{\lambda}^{\left(1\right)} & = & \frac{1}{16\pi^{2}}\left(5\lambda+6\frac{e^{4}}{\lambda}-6e^{2}\right)\left(\frac{2}{\varepsilon}-\gamma+\ln\left(4\pi\right)\right) \;. \label{zllt}
\end{eqnarray}
Before proceeding with the evaluation of $\Big(Z_{JJ},Z_{\Upsilon\Omega}^{\left(1\right)},Z_{\Omega\Omega}^{\left(1\right)}\Big)$, let us spend a few words on the relation (\ref{ea}), namely
\begin{equation} 
 e_0 A_{0\mu} = e A_\mu \;, \qquad Z_e Z_A^{1/2} = 1 \;. \label{ae22} 
\end{equation} 
There are several ways to test this relation computing, for example, the corrections to the three vertex $ve^2 h A_\mu A_\mu$. Moreover, the Higgs model offers a very nice and direct check of eq.(\ref{ae22}) through the corrections to the gauge boson mass. In fact, the above mentioned relation would imply that 
\begin{equation} 
\frac{1}{2} e^2_0 v^2_0 A_{0\mu} A_{0\mu} = \frac{Z_h}{2} e^2 v^2 A_\mu A_\mu \;, \label{check}
\end{equation}
meaning that the renormalization  factor of the gauge boson mass should be given entirely by the wave function renormalization of the Higgs field $h$, ${Z_h}$. That this is precisely the case follows from the evaluation of the two point gauge boson correlation function $\left\langle A_{\mu}\left(p\right)A_{\nu}\left(-p\right)\right\rangle^{\rm 1-loop}$ reported in Appendix \ref{B}, see equations (\ref{aa2p}),(\ref{zazh}).\\\\We now turn to the evaluation of $\Big(Z_{JJ},Z_{\Upsilon\Omega}^{\left(1\right)},Z_{\Omega\Omega}^{\left(1\right)}\Big)$, which can be extracted from the knowledge of the correlation functions $\left\langle h\left(x\right)O\left(y\right)\right\rangle $
and $\left\langle A_{\mu}\left(x\right)V_{\nu}\left(y\right)\right\rangle $, whose details are collected in Appendix \ref{C}. To that end we first evaluate $\left\langle h\left(x\right)\right\rangle _{J}$
and $\left\langle A_{\mu}\left(x\right)\right\rangle _{\Omega}$, where the use of the  indices  $J$ and $\Omega$ means that these sources are not yet set to zero, being treated as external fields in the Feynman rules, see Appendix \ref{A}. After the one-loop computation of  $\left\langle h\left(x\right)\right\rangle _{J}$
and $\left\langle A_{\mu}\left(x\right)\right\rangle _{\Omega}$, we differentiate them with respect to $J$ and $\Omega$, obtaining thus the desired correlation functions, {\it i.e.}
\begin{eqnarray}
\left.\frac{\delta\left\langle h\left(x\right)\right\rangle _{J}}{\delta J\left(y\right)}\right|_{J=\Omega=0} & = & \left\langle h\left(x\right)O\left(y\right)\right\rangle \;,  \nonumber \\
\left.\frac{\delta\left\langle A_{\mu}\left(x\right)\right\rangle _{\Omega}}{\delta\Omega_{\nu}\left(y\right)}\right|_{J=\Omega=0} & = & \left\langle A_{\mu}\left(x\right)V_{\nu}\left(y\right)\right\rangle \;. \label{des}
\end{eqnarray}
The first correction for $Z_{JJ}$ can be determined through the Green's
function $\left\langle h\left(x\right)O\left(y\right)\right\rangle $, whose calculation  can be found in Appendix \ref{C}: 
\begin{eqnarray}
Z_{JJ}^{\left(1\right)} & = & \frac{1}{16\pi^{2}}\left(2\lambda-3e^{2}\right)\left(\frac{2}{\varepsilon}-\gamma+\ln\left(4\pi\right)\right)\;. \label{zjj}
\end{eqnarray}
It is easy to see that (\ref{zjj}) agrees with the result given by the Ward identities and the algebraic analysis, eq.(\ref{zzz1}), {\it i.e.}
\begin{equation} 
Z_{JJ}^{\left(1\right)}=Z_{\lambda}^{\left(1\right)}+Z_{h}^{\left(1\right)}-2\frac{\left(\delta\sigma\right)^{\left(1\right)}_{\rm div}}{\lambda} \;. \label{zja}
\end{equation}
Another prediction of the Ward identities and the algebraic analysis is the remarkable result 
\begin{eqnarray}
Z_{\Omega\Omega} & = & 1 \;, \nonumber \\
Z_{\Upsilon\Omega} & = & - \frac{1}{2e}(Z_{A}- 1) \;. \label{rs1}
\end{eqnarray}
which means $Z_{\Omega\Omega}$ does not receive quantum corrections, while $Z_{\Upsilon\Omega}$ starts at  order $\hbar$ and can be expressed in terms of the gauge boson wave function renormalization factor $Z_{A}$. At one-loop order, we get 
\begin{eqnarray}
Z_{\Omega\Omega}^{\left(1\right)} & = & 0 \nonumber \;, \\
Z_{\Upsilon\Omega}^{\left(1\right)} & = & \frac{e}{96\pi^{2}}\left(\frac{2}{\varepsilon}-\gamma+\ln\left(4\pi\right)\right) \;. \label{rs2}
\end{eqnarray}
The explicit check of equations (\ref{rs2}) can be found in the last subsection of Appendix \ref{C}.

\section{Conclusion} 
In the present paper we have pursued the investigation started in \cite{Dudal:2019aew,Dudal:2019pyg} of the two BRST invariant local operators $(O,V_\mu)$, eq.(\ref{ovop}), by studying their renormalization properties, encoded in the renormalization of the corresponding external sources $(J, \Omega_\mu)$ needed to introduce them in the starting action $\Sigma$, eq.(\ref{cact}). As shown in \cite{Dudal:2019aew,Dudal:2019pyg}, these operators provide a BRST invariant framework to describe the Higgs particle and the gauge vector boson in the $U(1)$ Higgs model. As such, the current paper gives formal ground for the renormalization and subtraction procedures applied in \cite{Dudal:2019aew,Dudal:2019pyg} to obtain the spectral properties associated with the BRST-invariant, composite description of the physical degrees of freedom of the model. Moreover, the presented final bare action allows for perturbatively computing any renormalized n-point correlation function of the composite operators.\\\\The BRST invariant nature of $(O,V_\mu)$ has been exploited by making use of the Landau gauge condition, $\partial_\mu A_\mu=0$, which, due to the large set of Ward identities, provides several practical advantages with respect to the $R_\xi$ gauge as far as BRST invariant quantities are concerned. \\\\Our results are displayed in Section 5, where the explicit one-loop expression of the $Z$'s factors of both fields and operators have been displayed. We underline the fact that the vector operator $V_\mu$ mixes with $\partial_\nu F_{\nu \mu}$, a feature already observed in \cite{Clark:1974eq}. Moreover, a powerful Ward identity exists, eq.(\ref{Widom}), which has enabled us to obtain a purely algebraic characterization of the $2 \times 2$ mixing matrix. \\\\Our next goal will be that of looking at the Yang-Mills case, where a generalization of the operators $(O,V_\mu)$ can be constructed in order to have a BRST invariant setup for both Higgs and gauge vector bosons, see \cite{Maas:2019nso} and refs. therein for a general overview. This is the case, for instance, of the $SU(2)$ Yang-Mills model with a single Higgs field in the fundamental representation. Such a study could be the starting point to investigate the behaviour of  Yang-Mills-Higgs models in the infrared region, where possible non-perturbative effects related to the existence of  Gribov copies could be taken into account by means of the BRST invariant formulation of the Gribov-Zwanziger horizon function established in \cite{Capri:2015ixa}. These studies could lead to an interesting comparison with both present, see Ref. \cite{Maas:2019nso}, and future  lattice investigations of Yang-Mills-Higgs models.

\section*{Acknowledgements}
The authors would like to thank the Brazilian agencies CNPq and FAPERJ for financial support. This study was financed in part 
by the Coordena{\c c}{\~a}o de Aperfei{\c c}oamento de Pessoal de N{\'\i}vel Superior - Brasil (CAPES) - Finance Code 001. 
This paper is also part of the project INCT-FNA Process No. 464898/2014-5.
I.F. Justo acknowledges CAPES for the financial support under the project grant $88887.357904/2019-00$.
S.P. Sorella is a level $1$ CNPq researcher under the contract $301030/2019-7$. L.F. Palhares is a level $2$ CNPq researcher under contract $	311751/2019-9$. M.A.L. Capri is a level $2$ CNPq researcher under the contract $302040/2017-0$

\appendix

\section{Appendix A: propagators and one-loop Feynman rules \label{A}}

From the quadratic action, eq.(\ref{qq}), we can easily derive the following tree level  propagators

\begin{eqnarray}
\left\langle A_{\mu}\left(p\right)A_{\nu}\left(-p\right)\right\rangle _{\textrm{tree}} & = & \frac{1}{p^{2}+m^{2}}\Big(\delta_{\mu\nu} - \frac{p_\mu p_\nu}{p^2} \Big) \;,  \nonumber \\
\left\langle h\left(p\right)h\left(-p\right)\right\rangle _{\textrm{tree}} & = & \frac{1}{p^{2}+m_{h}^{2}}\;, \nonumber \\
\left\langle \rho\left(p\right)\rho\left(-p\right)\right\rangle _{\textrm{tree}} & = & \frac{1}{p^{2}}\;, \nonumber \\
\left\langle \overline{c}\left(p\right)c\left(-p\right)\right\rangle _{\textrm{tree}} & = & \frac{1}{p^{2}}\;, \nonumber \\
\left\langle A_{\mu}\left(p\right)b\left(-p\right)\right\rangle _{\textrm{tree}} & = & \frac{p_{\mu}}{p^{2}}\;, \nonumber \\
\left\langle \rho\left(p\right)b\left(-p\right)\right\rangle _{\textrm{tree}} & = & -i\frac{m^{2}}{p^{2}}\;. 
\label{treeprop} 
\end{eqnarray}
Let us also remind that the fields $(b,c,\overline{c})$ are non interacting fields, so that they will not enter into the interaction vertices. In order to give the Feynman rules, we shall make use of the following diagrammatic notation
\begin{eqnarray*}
\left\langle A_{\mu}\left(p\right)A_{\nu}\left(-p\right)\right\rangle _{\textrm{tree}} & = & \Diagram{g & g}
\;\;, \\
\left\langle h\left(p\right)h\left(-p\right)\right\rangle _{\textrm{tree}} & = & \Diagram{f & f}
\;\;, \\
\left\langle \rho\left(p\right)\rho\left(-p\right)\right\rangle _{\textrm{tree}} & = & \Diagram{h & h} \;\;.
\end{eqnarray*}  
The interaction action $S_{\rm I}$, eq.(\ref{eq:sinter}),   leads to the following Feynman rules, here given in  $d$-dimensional Euclidean space-time, with $d=4-\varepsilon$ and incomming momenta:

\begin{eqnarray}
\Diagram{\momentum{fA}{p}}
 & = & -(\delta \sigma)^{\left(1\right)}v^{3}\left(2\pi\right)^{d}\delta\left(p\right)-v\left(1+\left(Z_{JJ}^{\left(1\right)}+Z_{h}^{\left(1\right)}\right)\right)\widetilde{J}\left(p\right) \;, 
 \\
\Diagram{\momentum{hA}{p}}
 & = & -\frac{1}{2}\left(1+\left(Z_{\Omega}^{\left(1\right)}+Z_{h}^{\left(1\right)}\right)\right)vip_{\mu}\widetilde{\Omega}_{\mu}\left(p\right) \;, \\
\nonumber \\
\Diagram{\momentum{gA}{p}}
 & = & -\frac{1}{2}\left(1+\left(Z_{\Omega\Omega}^{\left(1\right)}+Z_{h}^{\left(1\right)}\right)\right)ev^{2}\widetilde{\Omega}_{\mu}\left(p\right)+Z_{\Upsilon\Omega}^{\left(1\right)}p^{2}P_{\mu\nu}\left(p\right)\widetilde{\Omega}_{\nu}\left(p\right) \;, \\
\nonumber \\
\Diagram{\momentum{gdA}{p} &  & \momentum{guV}{q}}
 & = & -\left(Z_{A}^{\left(1\right)}\left(p^{2}\delta_{\mu\nu}-p_{\mu}p_{\nu}\right)+Z_{h}^{\left(1\right)}m^{2}\delta_{\mu\nu}\right)\left(2\pi\right)^{d}\delta\left(p+q\right) \;, \\
\nonumber \\
\Diagram{\momentum{fdA}{p} &  & \momentum{fuV}{q}}
 & = & -\left(Z_{h}^{\left(1\right)}p^{2}+\left(Z_{\lambda}^{\left(1\right)}+ 2 Z_{h}^{\left(1\right)}\right)m_{h}^{2}+(\delta \sigma)^{\left(1\right)}v^{2}\right)\left(2\pi\right)^{d}\delta\left(p+q\right)\nonumber  \\
 &  & -\left(1+\left(Z_{JJ}^{\left(1\right)}+Z_{h}^{\left(1\right)}\right)\right)\widetilde{J}\left(p+q\right)\;, \\
\nonumber \\
\Diagram{\momentum{hdA}{p} &  & \momentum{huV}{q}}
 & = & -\left(Z_{h}^{\left(1\right)}p^{2}+(\delta \sigma)^{\left(1\right)}v^{2}\right)\left(2\pi\right)^{d}\delta\left(p+q\right)\nonumber   \\
 &  & -\left(1+\left(Z_{JJ}^{\left(1\right)}+Z_{h}^{\left(1\right)}\right)\right)\widetilde{J}\left(p+q\right)\;, \\
\nonumber \\
\Diagram{\momentum{gdA}{p} &  & \momentum{fuV}{q}}
 & = & -ev\left(1+\left(Z_{\Omega\Omega}^{\left(1\right)}+Z_{h}^{\left(1\right)}\right)\right)\widetilde{\Omega}_{\mu}\left(p+q\right)\;, \\
\nonumber \\
\Diagram{\momentum{gdA}{p} &  & \momentum{huV}{q}}
 & = & -ievZ_{h}^{\left(1\right)}q_{\mu}\left(2\pi\right)^{d}\delta\left(p+q\right)\;, \\
\nonumber \\
\Diagram{\momentum{fdA}{p} &  & \momentum{huV}{q}}
 & = & -\frac{1}{2}\left(1+\left(Z_{\Omega\Omega}^{\left(1\right)}+Z_{h}^{\left(1\right)}\right)\right)i\left(q_{\mu}-p_{\mu}\right)\widetilde{\Omega}_{\mu}\left(p+q\right) \;, \\
\nonumber \\
\Diagram{\momentum{hdA}{r} &  & \momentum{fuV}{q}\\
 & \momentum{gvA}{p}
}
 & = & ie\left(q_{\mu}-r_{\mu}\right)\left(1+Z_{h}^{\left(1\right)}\right)\left(2\pi\right)^{d}\delta\left(p+q+r\right)\;, \\
\nonumber \\
\Diagram{\momentum{fdA}{r} &  & \momentum{guV}{q}\\
 & \momentum{gvA}{p}
}
 & = & -2e^{2}v\left(1+Z_{h}^{\left(1\right)}\right)\delta_{\mu\nu}\left(2\pi\right)^{d}\delta\left(p+q+r\right)\;, \\
\nonumber \\
\Diagram{\momentum{gdA}{p} & \momentum{fuV}{r}\\
\momentum{guA}{q} & \momentum{fdV}{s}
}
 & = & -2e^{2}\left(1+Z_{h}^{\left(1\right)}\right)\delta_{\mu\nu}\left(2\pi\right)^{d}\delta\left(p+q+r+s\right) \;, \\
\nonumber \\
\Diagram{\momentum{gdA}{p} & \momentum{huV}{r}\\
\momentum{guA}{q} & \momentum{hdV}{s}
}
 & = & -2e^{2}\delta_{\mu\nu}\left(1+Z_{h}^{\left(1\right)}\right)\left(2\pi\right)^{d}\delta\left(p+q+r+s\right)\\
\nonumber \;, \\
\Diagram{\momentum{fdA}{r} &  & \momentum{fuV}{q}\\
 & \momentum{fvA}{p}
}
\Diagram{\\
\\
\\
\\
}
 & = & -3\lambda v\left(1+\left(Z_{\lambda}^{\left(1\right)}+2 Z_{h}^{\left(1\right)}\right)\right)\left(2\pi\right)^{d}\delta\left(p+q+r+s\right)\;, \\
\nonumber \\
\Diagram{\momentum{fdA}{p} & \momentum{fuV}{r}\\
\momentum{fuA}{q} & \momentum{fdV}{s}
}
 & = & -3\lambda\left(1+\left(Z_{\lambda}^{\left(1\right)}+2 Z_{h}^{\left(1\right)}\right)\right)\left(2\pi\right)^{d}\delta\left(p+q+r+s\right)\\
\nonumber \\
\Diagram{\momentum{hdA}{p} & \momentum{huV}{r}\\
\momentum{huA}{q} & \momentum{hdV}{s}
}
 & = & -3\lambda\left(1+\left(Z_{\lambda}^{\left(1\right)}+2 Z_{h}^{\left(1\right)}\right)\right)\left(2\pi\right)^{d}\delta\left(p+q+r+s\right) \;, \\
\nonumber \\
\Diagram{\momentum{fdA}{p} & \momentum{huV}{r}\\
\momentum{fuA}{q} & \momentum{hdV}{s}
}
 & = & -\lambda\left(1+\left(Z_{\lambda}^{\left(1\right)}+2 Z_{h}^{\left(1\right)}\right)\right)\left(2\pi\right)^{d}\delta\left(p+q+r+s\right) \;, \\
\nonumber \\
\Diagram{\momentum{hdA}{r} &  & \momentum{huV}{q}\\
 & \momentum{fvA}{p}
}
 & = & -\lambda v\left(1+\left(Z_{\lambda}^{\left(1\right)}+2 Z_{h}^{\left(1\right)}\right)\right)\left(2\pi\right)^{d}\delta\left(p+q+r\right) \;, \\
\nonumber \\
\Diagram{\momentum{fdA}{r} &  & \momentum{fuV}{q}\\
 & \momentum{gvA}{p}
}
 & = & -e\left(1+\left(Z_{\Omega\Omega}^{\left(1\right)}+Z_{h}^{\left(1\right)}\right)\right)\widetilde{\Omega}_{\mu}\left(p+q+r\right)\;, \\
\nonumber \\
\Diagram{\momentum{hdA}{r} &  & \momentum{huV}{q}\\
 & \momentum{gvA}{p}
}
 & = & -e\left(1+\left(Z_{\Omega\Omega}^{\left(1\right)}+Z_{h}^{\left(1\right)}\right)\right)\widetilde{\Omega}_{\mu}\left(p+q+r\right) \;, 
\end{eqnarray}
where $(\widetilde{J}\left(p\right), \widetilde{\Omega}_{\mu}\left(p\right))$ stand for the Fourier transformation of $(J(x), \Omega_\mu(x))$.

\section{Appendix B: Green's functions of the elementary fields \label{B} }

\subsection{The condition of vanishing tadpoles $\left\langle h\right\rangle =0$}
\begin{figure}[h]
\begin{center}
\includegraphics[width=.5\textwidth,angle=0]{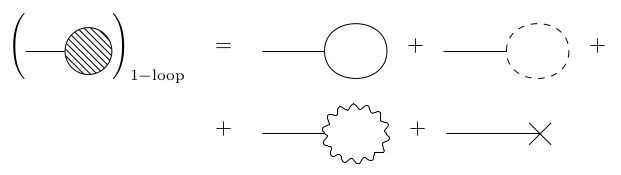}
\caption{One-loop diagrams contributing to the one-point Green's function $\langle h \rangle$  of the Higgs field $h$.
}
\label{1looptadpole}
\end{center}
\end{figure}

The diagrams contributing at one-loop order to $\left\langle h\right\rangle$ are displayed in  Figure \ref{1looptadpole}, where the counterterm $(\delta \sigma)^{(1)}$ has also been included. 

% \begin{eqnarray}
% \left( \feynmandiagram [baseline=(b.base), horizontal=a to b] {a --  b [blob],
% }; \right)_{\rm 1-loop} & = & \feynmandiagram [baseline=(b.base), horizontal=a to b] {
% a --  b
% -- [ half left] c, c --[half left] b,};+\feynmandiagram [baseline=(b.base), horizontal=a to b] {
% a --  b
% -- [ scalar, half left] c, c --[scalar, half left] b,}; \nonumber \\
% & & \feynmandiagram [baseline=(b.base), horizontal=a to b] {
% a --  b
% -- [photon, half left] c, c --[photon, half left] b,};+ 
% \feynmandiagram [baseline=(b.base), horizontal=a to b] {a --  [insertion=1] b } \;. \label{htadp}
% \end{eqnarray}
\noindent Using  dimensional regularization in the $\overline{\text{MS}}$ scheme, the evaluation of the first diagram gives: 
\begin{eqnarray*}
\left\langle h\right\rangle _{1}  =  \frac{1}{2}\frac{1}{m_{h}^{2}}\left(-3\lambda v\right)\int\frac{d^{d}k}{\left(2\pi\right)^{d}}\frac{1}{k^{2}+m_{h}^{2}}
  =  \frac{1}{m_{h}^{2}}\left(-\frac{3}{2}\lambda v\right)\chi\left(m_{h}^{2}\right) \;, \label{h1}
\end{eqnarray*}
where 
\begin{eqnarray}
\chi\left(M^{2}\right)  =  \int\frac{d^{d}k}{\left(2\pi\right)^{d}}\frac{1}{k^{2}+M^{2}}
  =  \frac{1}{\left(4\pi\right)^{\frac{d}{2}}}\Gamma\left(1-\frac{d}{2}\right)\left(M^{2}\right)^{\frac{d}{2}-1} \;. \label{chim}
\end{eqnarray}
For the other contributions, one gets sequentially
\begin{eqnarray*}
\left\langle h\right\rangle _{2}  =  \frac{1}{2}\frac{1}{m_{h}^{2}}\left(-\lambda v\right)\int\frac{d^{d}k}{\left(2\pi\right)}\frac{1}{k^{2}}  =  0 \;, \label{h2} 
\end{eqnarray*}

\begin{eqnarray*}
\left\langle h\right\rangle _{3}  =  \frac{1}{2}\frac{1}{m_{h}^{2}}\left(-2e^{2}\delta_{\mu\nu}\right)\int\frac{d^{d}k}{\left(2\pi\right)^{d}}\frac{1}{k^{2}+m^{2}}\left(\delta_{\mu\nu}-\frac{k_{\mu}k_{\nu}}{k^{2}}\right)
  =  \frac{1}{m_{h}^{2}}\left(-e^{2}v\right)\left(d-1\right)\chi\left(m^{2}\right) \;, \label{h3} 
\end{eqnarray*}

\begin{eqnarray*}
\left\langle h\right\rangle _{4} & = & \frac{1}{m_{h}^{2}}\left(-(\delta\sigma)^{\left(1\right)}v^{3}\right) \;. \label{h4}
\end{eqnarray*}
Imposing now the condition $\left\langle h\right\rangle =0$ at one-loop order, yields 
\begin{eqnarray}
(\delta\sigma)^{\left(1\right)}  =  \frac{1}{v^{2}}\left(-e^{2}\left(d-1\right)\chi\left(m^{2}\right)-\frac{3}{2}\lambda\chi\left(m_{h}^{2}
\right)\right) \;. \label{tdpcms}
\end{eqnarray}
Thus, for the one-loop divergent part of $(\delta\sigma)^{\left(1\right)}$ in the $\overline{\text{MS}}$ scheme, we have 
\begin{eqnarray}
 (\delta\sigma)^{\left(1\right)}_{\rm div} =  \frac{1}{\left(4\pi\right)^{2}}\frac{1}{v^{2}}\left(3e^{2}m^{2}+\frac{3}{2}\lambda m_{h}^{2}\right)
\left(\frac{2}{\varepsilon}-\gamma+\ln\left(4\pi\right)\right) \;. \label{divtad}
\end{eqnarray}

\subsection{The connected two-point function of the gauge field $\left\langle AA\right\rangle $}
\begin{figure}[h]
\begin{center}
\includegraphics[width=.7\textwidth,angle=0]{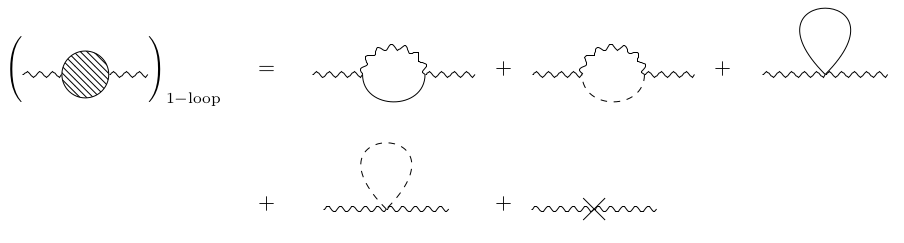}
\caption{
Diagrams contributing to the one-loop two-point Green's function of the Abelian vector field $A_{\mu}$.
}
\label{1loopphph}
\end{center}
\end{figure}

At one-loop order, for the connected two-point gauge field correlation function we have, including the counterterm, the diagrams shown in Figure \ref{1loopphph}.
% \begin{eqnarray*}
% \left( \feynmandiagram[layered layout,horizontal=a to c]{a -- [photon] b [blob], b -- [photon] c};\right)_{\rm 1-loop} & = & \feynmandiagram [layered layout, horizontal=b to c] {
% a -- [photon] b
% -- [photon, half left, looseness=1.5] c
% -- [ half left, looseness=1.5] b,
% c -- [photon] d,
% };  + \feynmandiagram [layered layout, horizontal=b to c] {
% a -- [photon] b
% -- [ half left, looseness=1.5] c
% -- [ scalar, half left, looseness=1.5] b,
% c -- [photon] d,};    \\
% & & + \feynmandiagram [layered layout, horizontal=a to b] {
% a -- [photon] b -- [out=135, in=45, loop, min distance=2cm] b,
% b -- [photon] c,};  
%  +  \feynmandiagram [layered layout, horizontal=a to b] {
% a -- [photon] b -- [scalar, out=135, in=45, loop, min distance=2cm] b,
% b -- [photon] c,}; 
%  +   \feynmandiagram[layered layout,horizontal=a to b]{a -- [photon,insertion=1] b,b --[photon] c,}. 
% \end{eqnarray*}

Defining
\begin{eqnarray*}
H\left(m_{1}^{2},m_{2}^{2},p^{2}\right) & = & p^{2}x\left(1-x\right)+xm_{1}^{2}+\left(1-x\right)m_{2}^{2}\;,
\end{eqnarray*}
for each diagram we obtain sequentially:
\begin{eqnarray*}
\left\langle A_{\mu}\left(p\right)A_{\nu}\left(-p\right)\right\rangle _{1}^{\rm 1-loop} & = & \frac{P_{\mu\alpha}\left(p\right)}{p^{2}+m^{2}}\frac{P_{\nu\beta}\left(p\right)}{p^{2}+m^{2}}\left(-2e^{2}v\delta_{\alpha\rho}\right)\left(-2e^{2}v\delta_{\beta\lambda}\right)\int\frac{d^{d}k}{\left(2\pi\right)^{d}}\frac{P_{\rho\lambda}\left(k\right)}{k^{2}+m^{2}}\frac{1}{\left(p-k\right)^{2}+m_{h}^{2}}\\
& = & \frac{P_{\mu\nu}\left(p\right)}{\left(p^{2}+m^{2}\right)^{2}}4e^{4}v^{2}\frac{1}{\left(4\pi\right)^{\frac{d}{2}}} \int_{0}^{1}dx \left[\Gamma\left(2-\frac{d}{2}\right)H\left(m^{2},m_{h}^{2},p^{2}\right)^{\frac{d}{2}-2}\right] \\
&  - & \frac{P_{\mu\nu}\left(p\right)}{\left(p^{2}+m^{2}\right)^{2}}4e^{4}v^{2}\frac{1}{\left(4\pi\right)^{\frac{d}{2}}} \int_{0}^{1} \left[\frac{1}{2m^{2}}\Gamma\left(1-\frac{d}{2}\right)\left(H\left(0,m_{h}^{2},p^{2}\right)^{\frac{d}{2}-1}-H\left(m^{2},m_{h}^{2},p^{2}\right)^{\frac{d}{2}-1}\right)\right] \;, \label{aa1c}
\end{eqnarray*}

\begin{eqnarray*}
\left\langle A_{\mu}\left(p\right)A_{\nu}\left(-p\right)\right\rangle _{2}^{\rm 1-loop} & = & \frac{P_{\mu\alpha}\left(p\right)}{p^{2}+m^{2}}\frac{P_{\nu\beta}\left(p\right)}{p^{2}+m^{2}}\int\frac{d^{d}k}{\left(2\pi\right)^{d}}ie\left(-2k_{\alpha}+p_{\alpha}\right)\frac{1}{k^{2}+m_{h}^{2}}ie\left(2k_{\beta}-p_{\beta}\right)\frac{1}{\left(p-k\right)^{2}}\\
 & = & \frac{P_{\mu\nu}\left(p\right)}{\left(p^{2}+m^{2}\right)^{2}}2e^{2}\frac{1}{\left(4\pi\right)^{\frac{d}{2}}}\int_{0}^{1}dx\Gamma\left(1-\frac{d}{2}\right)H\left(m_{h}^{2},0,p^{2}\right)^{\frac{d}{2}-1} \;, \label{aa2c}
\end{eqnarray*}

\begin{eqnarray*}
\left\langle A_{\mu}\left(p\right)A_{\nu}\left(-p\right)\right\rangle _{3}^{\rm 1-loop} & = & \frac{1}{2}\frac{P_{\mu\alpha}\left(p\right)}{p^{2}+m^{2}}\frac{P_{\nu\beta}\left(p\right)}{p^{2}+m^{2}}\left(-2e^{2}\delta_{\alpha\beta}\right)\int\frac{d^{d}k}{\left(2\pi\right)^{d}}\frac{1}{k^{2}+m_{h}^{2}}\\
 & = & \frac{P_{\mu\nu}\left(p\right)}{\left(p^{2}+m^{2}\right)^{2}}\left(-e^{2}\right)\frac{1}{\left(4\pi\right)^{\frac{d}{2}}}\Gamma\left(1-\frac{d}{2}\right)\left(m_{h}^{2}\right)^{\frac{d}{2}-1}\;, \label{aa3c}
\end{eqnarray*}

\begin{eqnarray}
\left\langle A_{\mu}\left(p\right)A_{\nu}\left(-p\right)\right\rangle _{4}^{\rm 1-loop} & = & \frac{1}{2}\frac{P_{\mu\alpha}\left(p\right)}{p^{2}+m^{2}}\frac{P_{\nu\beta}\left(p\right)}{p^{2}+m^{2}}\left(-2e^{2}\delta_{\alpha\beta}\right)\int\frac{d^{d}k}{\left(2\pi\right)^{d}}\frac{1}{k^{2}} 
  =  0 \;, \label{aa4c} 
\end{eqnarray}
where $P_{\mu\nu}(p)$ stands for the transverse projector 
\begin{equation} 
P_{\mu\nu}(p) = \left( \delta_{\mu\nu} - \frac{p_\mu p_\nu}{p^2}\right) \;. \label{tproj}
\end{equation}

For the one-loop counterterm, we get 
\begin{eqnarray}
\left\langle A_{\mu}\left(p\right)A_{\nu}\left(-p\right)\right\rangle^{\rm 1-loop} _{ct} & = & -\frac{P_{\mu\alpha}\left(p\right)}{p^{2}+m^{2}}\frac{P_{\nu\beta}\left(p\right)}{p^{2}+m^{2}}\left(Z_{A}^{\left(1\right)}\left(p^{2}\delta_{\alpha\beta}-p_{\alpha}p_{\beta}\right)+Z_{h}^{\left(1\right)}m^{2}\delta_{\alpha\beta}\right)\nonumber\\
 & = & -\frac{P_{\mu\nu}\left(p\right)}{\left(p^{2}+m^{2}\right)^{2}}\left(Z_{A}^{\left(1\right)}p^{2}+Z_{h}^{\left(1\right)}m^{2}\right) \;. \label{aact} 
\end{eqnarray}

Notice that, in the evaluation of $\left\langle A_{\mu}\left(p\right)A_{\nu}\left(-p\right)\right\rangle ^{\rm 1-loop}$,  we have not taken into account the contributions of tadpole diagrams which cancel by construction due to the condition $\left\langle h\right\rangle_{\rm 1-loop} =0$. Therefore, summing up all contributions, eqs.(\ref{aa1c})-(\ref{aact}), we get 
\begin{equation}
\left\langle A_{\mu}\left(p\right)A_{\nu}\left(-p\right)\right\rangle^{\rm 1-loop}   =  P_{\mu\nu}\left(p\right)
\Biggl(  \frac{1}{p^{2}+m^{2}}+\frac{\Pi_{AA}^{\rm 1-loop}}{\left(p^{2}+m^{2}\right)^{2}} \Biggl) \;, \label{faac} 
\end{equation}

\begin{eqnarray}
 \Pi_{AA}^{\rm 1-loop} & = & \left[4e^{4}v^{2}\frac{1}{\left(4\pi\right)^{\frac{d}{2}}}\int_{0}^{1}dx\left(\Gamma\left(2
 -\frac{d}{2}\right)H\left(m^{2},m_{h}^{2},p^{2}\right)^{\frac{d}{2}-2}+\frac{1}{2m^{2}}\Gamma\left(1-\frac{d}{2}\right)H\left(m^{2},m_{h}^{2},p^{2}\right)^{\frac{d}{2}-1}\right)\right. \nonumber \\
 &  & \left.-\frac{1}{\left(4\pi\right)^{\frac{d}{2}}}e^{2}\Gamma\left(1-\frac{d}{2}\right)\left(m_{h
 }^{2}\right)^{\frac{d}{2}-1}-\left(Z_{A}^{\left(1\right)}p^{2}+Z_{h}^{\left(1\right)}m^{2}\right)\right] \;. \label{piaa} 
\end{eqnarray}
From 
\begin{eqnarray*}
\Gamma\left(2-\frac{d}{2}\right) & = & \frac{2}{\varepsilon}-\gamma+O\left(\varepsilon\right) \;, \nonumber \\
\Gamma\left(1-\frac{d}{2}\right) & = & -\frac{2}{\varepsilon}+\gamma-1+O\left(\varepsilon\right) \;, \label{gammaexp}
\end{eqnarray*}
it follows that
\begin{eqnarray}
\left\langle A_{\mu}\left(p\right)A_{\nu}\left(-p\right)\right\rangle^{\rm 1-loop}  & = & P_{\mu\nu}\left(p\right)\left\{ \frac{1}{p^{2}+m^{2}}+\frac{1}{\left(p^{2}+m^{2}\right)^{2}}\right. \nonumber \\
 &  & \left[\frac{e^{2}}{\left(4\pi\right)^{2}}\left(\frac{2}{\varepsilon}-\gamma+\ln\left(4\pi\right)\right)\left(-\frac{p^{2}}{3}+3m^{2}\right)\right. \nonumber \\
 &  & -\left(Z_{A}^{\left(1\right)}p^{2}+Z_{h}^{\left(1\right)}m^{2}\right) \nonumber \\
 &  & -4e^{2}m^{2}\frac{1}{\left(4\pi\right)^{2}}\int_{0}^{1}dx\ln\left(\frac{H\left(m^{2},m_{h}^{2},p^{2}\right)}{\mu^{2}}\right) \nonumber \\
 &  & +2e^{2}\frac{1}{\left(4\pi\right)^{2}}\int_{0}^{1}dxH\left(m^{2},m_{h}^{2},p^{2}\right)\left(\ln\left(\frac{H\left(m^{2},m_{h}^{2},p^{2}\right)}{\mu^{2}}\right)-1\right)\nonumber \\
 &  & \left.\left.-\frac{1}{\left(4\pi\right)^{2}}e^{2}m_{h}^{2}\left(\ln\left(\frac{m_{h}^{2}}{\mu^{2}}\right)-1\right)\right]\right\} \;. \label{aa2p}
\end{eqnarray}
Thus, in the  $\overline{\text{MS}}$ scheme, for the one-loop renormalization factors $(Z_{A}^{\left(1\right)},Z_{h}^{\left(1\right)})$, we have 
\begin{eqnarray}
Z_{A}^{\left(1\right)} & = & -\frac{e^{2}}{48\pi^{2}}\left(\frac{2}{\varepsilon}-\gamma+\ln\left(4\pi\right)\right) \;, \nonumber \\
Z_{h}^{\left(1\right)} & = & \frac{3e^{2}}{16\pi^{2}}\left(\frac{2}{\varepsilon}-\gamma+\ln\left(4\pi\right)\right) \;, \label{zazh}
\end{eqnarray}
so that the one-loop renormalized connected two point gauge correlation function turns out to be 
\begin{equation} 
\left\langle A_{\mu}\left(p\right)A_{\nu}\left(-p\right)\right\rangle^{\rm 1-loop}_{\rm ren}   =  P_{\mu\nu}\left(p\right)\Biggl( \frac{1}{p^{2}+m^{2}}+\frac{(\Pi_{AA}^{\rm 1-loop})_{\rm ren}}{\left(p^{2}+m^{2}\right)^{2}}\Biggl) \;, \label{piaaren} 
\end{equation} 
with 
\begin{eqnarray}
(\Pi_{AA}^{\rm 1-loop})_{\rm ren} & = & \frac{e^{2}}{\left(4\pi\right)^{2}}
 \left[-4m^{2}\int_{0}^{1}dx\ln\left(\frac{H\left(m^{2},m_{h}^{2},p^{2}\right)}{\mu^{2}}\right)+2\int_{0}^{1}dxH\left(m^{2},m_{h}^{2}\right)\left(\ln\left(\frac{H\left(m^{2},m_{h}^{2},p^{2}\right)}{\mu^{2}}\right)-1\right)\right] \nonumber \\
 & - & \frac{e^{2}}{\left(4\pi\right)^{2}}\left[m_{h}^{2}\left(\ln\left(\frac{m_{h}^{2}}{\mu^{2}}\right)-1\right)\right] \;. \label{aarenfinal}
\end{eqnarray}
Let us end this subsection by underlining that, from eqs.(\ref{aa2p}),(\ref{zazh}), it follows that the renormalization factor of the gauge boson mass is given by the wave function $Z_h$ of the Higgs field, as stated by the Ward identity (\ref{ea}).

\subsection{The connected one-loop two-point correlation function of the Higgs field $\left\langle hh\right\rangle $}
\begin{figure}[h]
\begin{center}
\includegraphics[width=.8\textwidth,angle=0]{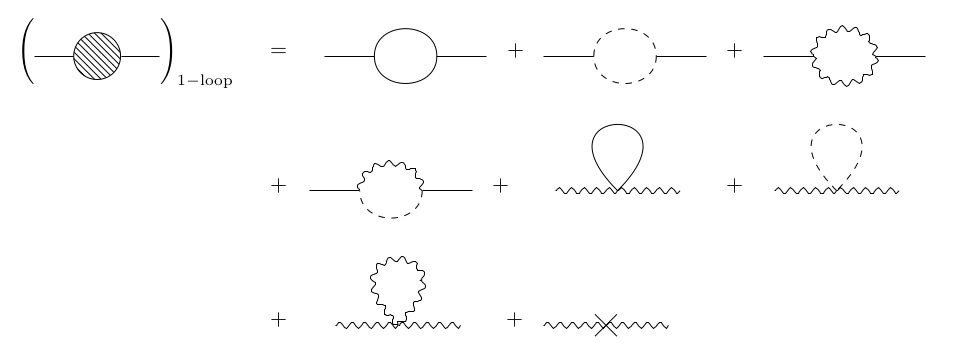}
\caption{
Feynman diagrams which contribute to the one-loop two-point function of the Higgs field $h$.
}
\label{1loophh}
\end{center}
\end{figure}

Let us now consider the connected one-loop two point function of the Higgs field, whose  contributing Feynman diagrams are depicted in Figure \ref{1loophh}. 
% \begin{eqnarray}
% \left( \feynmandiagram[layered layout,horizontal=a to c]{a -- b [blob], b -- c}; \right)_{\rm 1-loop}& = & \feynmandiagram [layered layout, horizontal=b to c] {
% a -- b
% -- [ half left, looseness=1.5] c
% -- [ half left, looseness=1.5] b,
% c -- d,
% };  + \feynmandiagram [layered layout, horizontal=b to c] {
% a -- b
% -- [scalar, half left, looseness=1.5] c
% -- [scalar, half left, looseness=1.5] b,
% c -- d,
% }; \nonumber \\
% & & \feynmandiagram [layered layout, horizontal=b to c] {
% a -- b
% -- [photon, half left, looseness=1.5] c
% -- [photon, half left, looseness=1.5] b,
% c -- d,
% };  + \feynmandiagram [layered layout, horizontal=b to c] {
% a -- b
% -- [photon, half left, looseness=1.5] c
% -- [scalar, half left, looseness=1.5] b,
% c -- d,
% }; \nonumber \\
% & & + \feynmandiagram [layered layout, horizontal=a to b] {
% a -- b -- [out=135, in=45, loop, min distance=2cm] b,
% b -- c,};  + \feynmandiagram [layered layout, horizontal=a to b] {
% a -- b -- [scalar, out=135, in=45, loop, min distance=2cm] b,
% b -- c,};  +  \feynmandiagram [layered layout, horizontal=a to b] {
% a -- b -- [photon, out=135, in=45, loop, min distance=2cm] b,
% b -- c,}; \nonumber \\
% & & + \feynmandiagram[layered layout,horizontal=a to b]{a -- [insertion=1] b, b -- c}; 
% \end{eqnarray}

As done before in the evaluation of $\left\langle A_{\mu}\left(p\right)A_{\nu}\left(-p\right)\right\rangle ^{\rm 1-loop}$,  we have not taken into account  contributions of tadpole diagrams which vanish  due to the condition $\left\langle h\right\rangle_{\rm 1-loop} =0$.\\\\For each single contribution we get 
\begin{eqnarray}
\left\langle h\left(p\right)h\left(-p\right)\right\rangle _{1}^{\rm 1-loop} & = & \frac{1}{2}\frac{1}{\left(p^{2}+m_{h}^{2}\right)^{2}}\left(-3v\lambda\right)^{2}\int\frac{d^{d}k}{\left(2\pi\right)^{d}}\frac{1}{k^{2}+m_{h}^{2}}\frac{1}{\left(p-k\right)^{2}+m_{h}^{2}} \nonumber \\
 & = & \frac{1}{\left(p^{2}+m_{h}^{2}\right)^{2}}\frac{9}{2}m_{h}^{2}\frac{\lambda}{\left(4\pi\right)^{\frac{d}{2}}}\Gamma\left(2-\frac{d}{2}\right)\int_{0}^{1}dxH\left(m_{h}^{2},m_{h}^{2},p^{2}\right)^{\frac{d}{2}-2}\nonumber \\
 & = & \frac{1}{\left(p^{2}+m_{h}^{2}\right)^{2}}\left[\frac{9}{2}m_{h}^{2}\frac{\lambda}{\left(4\pi\right)^{2}}\left(\frac{2}{\varepsilon}-\gamma+\ln\left(4\pi\right)\right)-\frac{9}{2}m_{h}^{2}\frac{\lambda}{\left(4\pi\right)^{2}}\int_{0}^{1}dx\ln\left(\frac{H\left(m_{h}^{2},m_{h}^{2},p^{2}\right)}{\mu^{2}}\right)\right] \nonumber \\[3mm] \label{hhc1} 
\end{eqnarray}

\begin{eqnarray}
\left\langle h\left(p\right)h\left(-p\right)\right\rangle _{2}^{\rm 1-loop} & = & \frac{1}{2}\frac{1}{\left(p^{2}+m_{h}^{2}\right)^{2}}\left(-v\lambda\right)^{2}\int\frac{d^{d}k}{\left(2\pi\right)^{d}}\frac{1}{k^{2}}\frac{1}{\left(p-k\right)^{2}}\nonumber \\
 & = & \frac{1}{\left(p^{2}+m_{h}^{2}\right)^{2}}\frac{1}{2}v^{2}\lambda^{2}\frac{1}{\left(4\pi\right)^{\frac{d}{2}}}\Gamma\left(2-\frac{d}{2}\right)\int_{0}^{1}dxH\left(0,0,p^{2}\right)^{\frac{d}{2}-2} \nonumber \\
 & = & \frac{1}{\left(p^{2}+m_{h}^{2}\right)^{2}}\left[\frac{1}{2}m_{h}^{2}\frac{\lambda}{\left(4\pi\right)^{2}}\left(\frac{2}{\varepsilon}-\gamma+\ln\left(4\pi\right)\right)-\frac{1}{2}m_{h}^{2}\frac{\lambda}{\left(4\pi\right)^{2}}\int_{0}^{1}dx\ln\left(\frac{H\left(0,0,p^{2}\right)}{\mu^{2}}\right)\right] \nonumber 
 \\[3mm] \label{hhc2}
\end{eqnarray}

\begin{equation}
  \left\langle h(p) h(-p)\right\rangle _{3}^{1} = \frac{\Pi_3^{1}}{(p^2 + m^2_h)^2} \;, \label{pi3hh}
\end{equation} 
with 

\begin{eqnarray}
 \Pi_3^{1} &  = & \frac{1}{2}\left(-2e^{2}v\delta_{\mu\alpha}\right)\left(-2e^{2}v\delta_{\nu\beta}\right)\int\frac{d^{d}k}{\left(2\pi\right)^{d}}\frac{P_{\mu\nu}\left(k\right)}{k^{2}+m^{2}}\frac{P_{\alpha\beta}\left(p-k\right)}{\left(p-k\right)^{2}+m^{2}} \nonumber \\
 & = & \frac{(2e^{4} v^{2})}{\left(4\pi\right)^{\frac{d}{2}}}\int_{0}^{1}dx\left\{ \left(d-1\right)\Gamma\left(2-\frac{d}{2}\right)H\left(m^{2},m^{2},p^{2}\right)^{\frac{d}{2}-2}\right. \nonumber\\
 &  & -\Gamma\left(2-\frac{d}{2}\right)\frac{p^{2}}{m^{2}}\left(H\left(m^{2},0,p^{2}\right)^{\frac{d}{2}-2}-H\left(m^{2},m^{2},p^{2}\right)^{\frac{d}{2}-2}\right) \nonumber \\
 &  & +\Gamma\left(2-\frac{d}{2}\right)\frac{p^{4}}{m^{4}}\left(1-x\right)^{2}\left(H\left(0,0,p^{2}\right)^{\frac{d}{2}-2}-H\left(m^{2},0,p^{2}\right)^{\frac{d}{2}-2}-H\left(0,m^{2},p^{2}\right)^{\frac{d}{2}-2}+H\left(m^{2},m^{2},p^{2}\right)^{\frac{d}{2}-2}\right)\nonumber \\
 &  & \left.+\Gamma\left(1-\frac{d}{2}\right)\frac{p^{2}}{2m^{4}}\left(H\left(0,0,p^{2}\right)^{\frac{d}{2}-1}-2H\left(m^{2},0,p^{2}\right)^{\frac{d}{2}-1}+H\left(m^{2},m^{2},p^{2}\right)^{\frac{d}{2}-1}\right)\right\} \nonumber \\
 & = & \left\{ 6m^{2}\frac{e^{2}}{\left(4\pi\right)^{2}}\left(\frac{2}{\varepsilon}-\gamma+\ln\left(4\pi\right)\right)\right. \nonumber \\
 & + & \frac{ e^2m^2}{\left(8\pi^2\right)}\int_{0}^{1}dx\left[-2-3\ln\left(\frac{H\left(m^{2},m^{2},p^{2}\right)}{\mu^{2}}\right)+\frac{p^{2}}{m^{2}}\left(\ln\left(\frac{H\left(m^{2},0,p^{2}\right)}{\mu^{2}}\right)-\ln\left(\frac{H\left(m^{2},m^{2},p^{2}\right)}{\mu^{2}}\right)\right)\right. \nonumber \\
 & - & \frac{p^{4}}{m^{4}}  \left(1-x\right)^{2}\left(\ln\left(\frac{H\left(0,0,p^{2}\right)}{\mu^{2}}\right)-\ln\left(\frac{H\left(m^{2},0,p^{2}\right)}{\mu^{2}}\right)-\ln\left(\frac{H\left(0,m^{2},p^{2}\right)}{\mu^{2}}\right)+\ln\left(\frac{H\left(m^{2},m^{2},p^{2}\right)}{\mu^{2}}\right)\right) \nonumber \\
 & + & \left. \left. \frac{p^{2}}{2m^{4}}\left(H\left(0,0,p^{2}\right)\ln\left(\frac{H\left(0,0,p^{2}\right)}{\mu^{2}}\right)-2H\left(m^{2},0,p^{2}\right)\ln\left(\frac{H\left(m^{2},0,p^{2}\right)}{\mu^{2}}\right)+H\left(m^{2},m^{2},p^{2}\right)\ln\left(\frac{H\left(m^{2},m^{2},p^{2}\right)}{\mu^{2}}\right)\right)\right]\right\} \nonumber \\[3mm]
 \label{hhc3}
\end{eqnarray}

\begin{eqnarray}
\left\langle h\left(p\right)h\left(-p\right)\right\rangle _{4}^{\rm 1-loop} & = & \frac{1}{\left(p^{2}+m_{h}^{2}\right)^{2}}\int\frac{d^{d}k}{\left(2\pi\right)^{d}}ie\left(2p_{\mu}-k_{\mu}\right)\frac{P_{\mu\nu}\left(k\right)}{k^{2}+m^{2}}ie\left(-2p_{\nu}+k_{\nu}\right)\frac{1}{\left(p-k\right)^{2}} \nonumber \\
 & = & \frac{1}{\left(p^{2}+m_{h}^{2}\right)^{2}}4e^{2}\frac{1}{\left(4\pi\right)^{\frac{d}{2}}}\int_{0}^{1}dx\left\{ p^{2}\Gamma\left(2-\frac{d}{2}\right)H\left(m^{2},0,p^{2}\right)^{\frac{d}{2}-2}\right. \nonumber \\
 &  & -\frac{p^{4}}{m^{2}}\left(1-x\right)^{2}\Gamma\left(2-\frac{d}{2}\right)\left(H\left(0,0,p^{2}\right)^{\frac{d}{2}-2}-H\left(m^{2},0,p^{2}\right)^{\frac{d}{2}-2}\right) \nonumber \\
 &  & \left.-\frac{p^{2}}{2m^{2}}\Gamma\left(1-\frac{d}{2}\right)\left(H\left(0,0,p^{2}\right)^{\frac{d}{2}-1}-H\left(m^{2},0,p^{2}\right)^{\frac{d}{2}-1}\right)\right\} \nonumber  \\
 & = & \frac{1}{\left(p^{2}+m_{h}^{2}\right)^{2}}\left\{ 3p^{2}\frac{e^{2}}{\left(4\pi\right)^{2}}\left(\frac{2}{\varepsilon}-\gamma+\ln\left(4\pi\right)\right)\right. \nonumber \\
 &  & 4e^{2}\frac{1}{\left(4\pi\right)^{2}}\int_{0}^{1}dx\left[-p^{2}\ln\left(\frac{H\left(m^{2},0,p^{2}\right)}{\mu^{2}}\right)+\frac{p^{4}}{m^{2}}\left(1-x\right)^{2}\left(\ln\left(\frac{H\left(0,0,p^{2}\right)}{\mu^{2}}\right)-\ln\left(\frac{H\left(m^{2},0,p^{2}\right)}{\mu^{2}}\right)\right)\right. \nonumber \\
 &  & \left.\left.-\frac{p^{2}}{2m^{2}}\left(H\left(0,0,p^{2}\right)\left(\ln\left(\frac{H\left(0,0,p^{2}\right)}{\mu^{2}}\right)-1\right)-H\left(m^{2},0,p^{2}\right)\left(\ln\left(\frac{H\left(m^{2},0,p^{2}\right)}{\mu^{2}}\right)-1\right)\right)\right]\right\} \nonumber \\[3mm] \label{hhc4}
\end{eqnarray}

\begin{eqnarray}
\left\langle h\left(p\right)h\left(-p\right)\right\rangle _{5}^{\rm 1-loop} & = & \frac{1}{2}\frac{1}{\left(p^{2}+m_{h}^{2}\right)^{2}}\left(-3\lambda\right)\int\frac{d^{d}k}{\left(2\pi\right)^{d}}\frac{1}{k^{2}+m_{h}^{2}} \nonumber \\
 & = & \frac{1}{\left(p^{2}+m_{h}^{2}\right)^{2}}\left[-\frac{3}{2}\lambda\frac{1}{\left(4\pi\right)^{\frac{d}{2}}}\Gamma\left(1-\frac{d}{2}\right)\left(m_{h}^{2}\right)^{\frac{d}{2}-1}\right] \nonumber \\
 & = & \frac{1}{\left(p^{2}+m_{h}^{2}\right)^{2}}\left[\frac{3}{2}m_{h}^{2}\frac{\lambda}{\left(4\pi\right)^{2}}\left(\frac{2}{\varepsilon}-\gamma+\ln\left(4\pi\right)\right)-\frac{3}{2}\frac{\lambda}{\left(4\pi\right)^{2}}m_{h}^{2}\left(\ln\left(\frac{m_{h}^{2}}{\mu^{2}}\right)-1\right)\right] \label{hhc5}
\end{eqnarray}
\\[4mm]
\begin{equation}
\left\langle h\left(p\right)h\left(-p\right)\right\rangle _{6}^{\rm 1-loop}  =  \frac{1}{2}\frac{1}{\left(p^{2}+m_{h}^{2}\right)^{2}}\left(-\lambda\right)\int\frac{d^{d}k}{\left(2\pi\right)^{d}}\frac{1}{k^{2}}
  =  0 \label{hhc6}
\end{equation}
\\[4mm]
\begin{eqnarray}
\left\langle h\left(p\right)h\left(-p\right)\right\rangle _{7}^{\rm 1-loop} & = & \frac{1}{2}\frac{1}{\left(p^{2}+m_{h}^{2}\right)^{2}}\left(-2e^{2}\delta_{\mu\nu}\right)\int\frac{d^{d}k}{\left(2\pi\right)^{d}}\frac{P_{\mu\nu}\left(k\right)}{k^{2}+m^{2}} \nonumber \\
 & = & \frac{1}{\left(p^{2}+m_{h}^{2}\right)^{2}}\left[-e^{2}\left(d-1\right)\frac{1}{\left(4\pi\right)^{\frac{d}{2}}}\Gamma\left(1-\frac{d}{2}\right)\left(m^{2}\right)^{\frac{d}{2}-1}\right] \nonumber \\
 & = & \frac{1}{\left(p^{2}+m_{h}^{2}\right)^{2}}\left[3m^{2}\frac{e^{2}}{\left(4\pi\right)^{2}}\left(\frac{2}{\varepsilon}-\gamma+\ln\left(4\pi\right)\right)-e^{2}\frac{1}{\left(4\pi\right)^{2}}m^{2}\left(3\ln\left(\frac{m^{2}}{\mu^{2}}\right)-1\right)\right] \label{hhc7}
\end{eqnarray}
Finally, for the one-loop counterterm, we have 
\begin{eqnarray}
\left\langle h\left(p\right)h\left(-p\right)\right\rangle _{ct}^{\rm 1-loop}  =  \frac{1}{\left(p^{2}+m_{h}^{2}\right)^{2}}\left[-\left(Z_{h}^{\left(1\right)}p^{2}+\left(Z_{\lambda}^{\left(1\right)}+Z_{h}^{\left(1\right)}\right)m_{h}^{2}+(\delta \sigma)^{\left(1\right)}v^{2}\right)\right]
\end{eqnarray}
so that 
\begin{eqnarray}
Z_{h}^{\left(1\right)} & = & \frac{3e^{2}}{16\pi^{2}}\left(\frac{2}{\varepsilon}-\gamma+\ln\left(4\pi\right)\right)\;, \nonumber \\
Z_{\lambda}^{\left(1\right)} & = & \frac{1}{16\pi^{2}}\left(5\lambda+6\frac{e^{4}}{\lambda}-6e^{2}\right)\left(\frac{2}{\varepsilon}-\gamma+\ln\left(4\pi\right)\right) \;. \label{zhhf}
\end{eqnarray}
It is worth remarking that the value of $Z_{h}^{\left(1\right)}$ in eq.(\ref{zhhf}) is in full agreement, as expected, with that obtained from the evaluation of $\left\langle A_{\mu}\left(p\right)A_{\nu}\left(-p\right)\right\rangle^{\rm 1-loop}_{\rm ren}$, see eq.(\ref{zazh}). As already mentioned, this feature offers a very simple check of the nonrenormalization property $e_0 A_{0\mu} = e A_\mu$ in the Abelian Higgs model.

\subsection{The connected one-loop two-point correlation function of the Goldstone field $\left\langle \rho\rho\right\rangle $}
\begin{figure}[h]
\begin{center}
\includegraphics[width=.8\textwidth,angle=0]{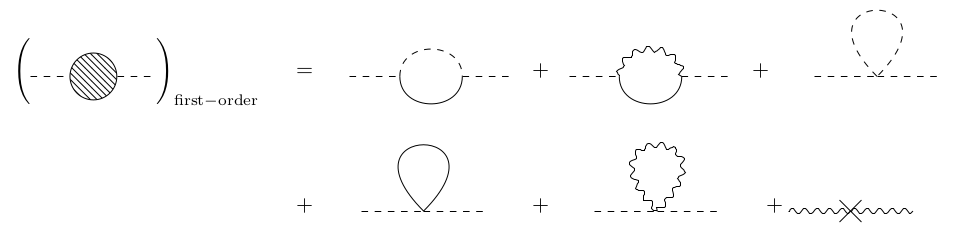}
\caption{
Feynman diagrams contributing to the two-point Green's function of the Goldstone field $\rho$.
}
\label{goldstgoldst}
\end{center}
\end{figure}

This subsection is devoted to the study of the {\it would-be Goldstone boson} two-point correlation function $\left\langle \rho\rho\right\rangle $. This will provide an additional check of the Ward identities displayed by the $U(1)$ Higgs model. In particular, in the Landau gauge, the field $\rho$ ought  to be massless as a consequence of the global Ward identity, eq.(\ref{gboW}), which, when written on the generator $\Gamma$ of the $1PI$ connected Green functions, takes the form 
\begin{equation} 
\int d^{4}x\left[-\rho\frac{\delta\Gamma}{\delta h}+\left(v+h\right)\frac{\delta\Gamma}{\delta\rho}-R\frac{\delta\Gamma}{\delta L}+L\frac{\delta\Gamma}{\delta R}\right] = 0\label{eq:global} \;. \label{gboWq}
\end{equation} 
Indeed, acting now on eq.(\ref{gboWq}) with the test operator $\frac{\delta}{\delta \rho(y)}$, using the tadpole condition $\langle h \rangle=0$, setting all sources and fields to zero and taking the Fourier transform one gets, in momentum space, 
\begin{equation} 
\Gamma(p^2=0)_{\rho\rho} = 0 \;, \label{mrho} 
\end{equation}
showing in fact that the Goldstone mode $\rho$ remains massless in the Landau gauge (see also  its tree level propagator in eq.(\ref{treeprop})). \\\\Let us confirm here this property by an explicit one-loop calculation, amounting to compute the  diagrams depicted in Figure \ref{goldstgoldst}.
% \begin{eqnarray}
% \left( \feynmandiagram[layered layout,horizontal=a to c]{a --[scalar] b [blob], b --[scalar] c}; \right)_{\rm first-order}& = & \feynmandiagram [layered layout, horizontal=b to c] {
% a --[scalar] b
% -- [scalar, half left, looseness=1.5] c
% -- [ half left, looseness=1.5] b,
% c --[scalar] d,
% };  + \feynmandiagram [layered layout, horizontal=b to c] {
% a --[scalar] b
% -- [photon, half left, looseness=1.5] c
% -- [ half left, looseness=1.5] b,
% c --[scalar] d,
% }; \nonumber \\
% & & + \feynmandiagram [layered layout, horizontal=a to b] {
% a --[scalar] b -- [scalar, out=135, in=45, loop, min distance=2cm] b,
% b --[scalar] c,};  + \feynmandiagram [layered layout, horizontal=a to b] {
% a --[scalar] b -- [out=135, in=45, loop, min distance=2cm] b,
% b --[scalar] c,};  +  \feynmandiagram [layered layout, horizontal=a to b] {
% a --[scalar] b -- [photon, out=135, in=45, loop, min distance=2cm] b,
% b --[scalar] c,}; \nonumber \\
% & & + \feynmandiagram[layered layout,horizontal=a to b]{a -- [scalar, insertion=1] b, b --[scalar] c}; 
% \end{eqnarray}

For each contribution we have 
\begin{eqnarray}
\left\langle \rho \rho\right\rangle _{1}^{\rm 1-loop}(p^2) & = & \frac{1}{p^{4}}\left(-\lambda v\right)^{2}\int\frac{d^{d}k}{\left(2\pi\right)^{d}}\frac{1}{k^{2}}\frac{1}{\left(p-k\right)^{2}+m_{h}^{2}} \nonumber \\
 & = & \frac{1}{p^{4}}\left(\lambda^{2}v^{2}\right)\frac{1}{\left(4\pi\right)^{\frac{d}{2}}}\Gamma\left(2-\frac{d}{2}\right)\int_{0}^{1}dxH\left(0,m_{h}^{2},p^{2}\right)^{\frac{d}{2}-2} \nonumber \\
 & = & \frac{1}{p^{4}}\left(\lambda m_{h}^{2}\right)\frac{1}{\left(4\pi\right)^{2}}\int_{0}^{1}dx\left(\frac{2}{\varepsilon}-\gamma+\ln\left(4\pi\right)-\ln\left(\frac{H\left(0,m_{h}^{2},p^{2}\right)}{\mu^{2}}\right)\right)\nonumber \\
 & = & \frac{1}{p^{4}}\left(\lambda m_{h}^{2}\right)\frac{1}{\left(4\pi\right)^{2}}\int_{0}^{1}dx\left(\frac{2}{\varepsilon}-\gamma+\ln\left(4\pi\right)-\ln\left(\frac{H\left(0,m_{h}^{2},p^{2}\right)}{\mu^{2}}\right)\right) \;, \label{rhorho1}
\end{eqnarray}

\begin{eqnarray}
\left\langle \rho \rho \right\rangle _{2}^{\rm 1-loop}(p^2) & = & \frac{1}{p^{4}}\int\frac{d^{d}k}{\left(2\pi\right)^{d}}\frac{P_{\mu\nu}\left(k\right)}{k^{2}+m^{2}}ie\left(-2p_{\mu}+k_{\mu}\right)\frac{1}{\left(p-k\right)^{2}+m_{h}^{2}}ie\left(2p_{\nu}-k_{\nu}\right) \nonumber \\
 & = & \frac{1}{p^{4}}4e^{2}\frac{1}{\left(4\pi\right)^{\frac{d}{2}}}\int_{0}^{1}dx\left\{ p^{2}\Gamma\left(2-\frac{d}{2}\right)H\left(m^{2},m_{h}^{2},p^{2}\right)^{\frac{d}{2}-2}\right. \nonumber \\
 &  & -\frac{p^{4}}{m^{2}}\left(1-x\right)^{2}\Gamma\left(2-\frac{d}{2}\right)\left(H\left(0,m_{h}^{2},p^{2}\right)^{\frac{d}{2}-2}-H\left(m^{2},m_{h}^{2},p^{2}\right)^{\frac{d}{2}-2}\right) \nonumber \\
 &  & \left.-\frac{p^{2}}{2m^{2}}\Gamma\left(1-\frac{d}{2}\right)\left(H\left(0,m_{h}^{2},p^{2}\right)^{\frac{d}{2}-1}-H\left(m^{2},m_{h}^{2},p^{2}\right)^{\frac{d}{2}-1}\right)\right\} \nonumber \\
 & = & \frac{1}{p^{4}}\left\{ \frac{e^{2}}{\left(4\pi\right)^{2}}3p^{2}\left(\frac{2}{\varepsilon}-\gamma+\ln\left(4\pi\right)\right)\right. \nonumber \\
 && 
 +4\frac{e^{2}}{\left(4\pi\right)^{2}}\int_{0}^{1}dx\left[-p^{2}\ln\left(\frac{H\left(m^{2},m_{h}^{2},p^{2}\right)}{\mu^{2}}\right)
 \right.
 \nonumber \\
 &&
 +\frac{p^{4}}{m^{2}}\left(1-x\right)^{2}\left(\ln\left(\frac{H\left(0,m_{h}^{2},p^{2}\right)}{\mu^{2}}\right)-\ln\left(\frac{H\left(m^{2},m_{h}^{2},p^{2}\right)}{\mu^{2}}\right)\right)
 \nonumber \\
 && 
 -\frac{p^{2}}{2m^{2}}\left(H\left(0,m_{h}^{2},p^{2}\right)\left(\ln\left(\frac{H\left(0,m_{h},p^{2}\right)}{\mu^{2}}\right)-1\right)
 \right.
 \nonumber \\
 &&
 \left.\left.\left.
 -H\left(m^{2},m_{h}^{2},p^{2}\right)\left(\ln\left(\frac{H\left(m^{2},m_{h}^{2},p^{2}\right)}{\mu^{2}}\right)-1\right)\right)\right]\right\} \;, 
 \label{rhorho2}
\end{eqnarray}

\begin{eqnarray}
\left\langle \rho\left(p\right)\rho\left(-p\right)\right\rangle _{3}^{\rm 1-loop}(p^2)  =  \frac{1}{2}\frac{1}{p^{4}}\left(-3\lambda\right)\int\frac{d^{d}k}{\left(2\pi\right)^{d}}\frac{1}{k^{2}}
  =  0 \;, \label{rhorho3}
\end{eqnarray}

\begin{eqnarray}
\left\langle \rho\left(p\right)\rho\left(-p\right)\right\rangle _{4}^{\rm 1-loop}(p^2)  =  \frac{1}{2}\frac{1}{p^{4}}\left(-\lambda\right)\int\frac{d^{d}k}{\left(2\pi\right)^{d}}\frac{1}{k^{2}+m_{h}^{2}}
  =  \frac{1}{p^{4}}\left[-\frac{\lambda}{2}\frac{1}{\left(4\pi\right)^{\frac{d}{2}}}\Gamma\left(1-\frac{d}{2}\right)\left(m_{h}^{2}\right)^{\frac{d}{2}-1}\right] \;, \label{rhorho4}
\end{eqnarray}

\begin{eqnarray}
\left\langle \rho\left(p\right)\rho\left(-p\right)\right\rangle _{5}^{1-loop} & = & \frac{1}{2}\frac{1}{p^{4}}\left(-2e^{2}\delta_{\mu\nu}\right)\int\frac{d^{d}k}{\left(2\pi\right)^{d}}\frac{P_{\mu\nu}\left(k\right)}{k^{2}+m^{2}} \nonumber \\
 & = & \frac{1}{p^{4}}\left[-e^{2}\left(d-1\right)\frac{1}{\left(4\pi\right)^{\frac{d}{2}}}\Gamma\left(1-\frac{d}{2}\right)\left(m^{2}\right)^{\frac{d}{2}-1}\right] \;. \label{rhorho5}
\end{eqnarray}
For the one-loop counterterm we have 
\begin{eqnarray}
\left\langle \rho\rho\right\rangle _{ct}^{\rm 1-loop}(p^2) & = & \frac{1}{p^{4}}\left[-\left(Z_{h}^{\left(1\right)}p^{2}+(\delta \sigma)^{\left(1\right)}v^{2}\right)\right] \;. \label{rhorhoct}
\end{eqnarray}
Since

\begin{eqnarray}
\left\langle \rho\rho\right\rangle _{div}^{\rm 1-loop}(p^2) & = & \frac{1}{p^{4}}\frac{1}{\left(4\pi\right)^{2}}\left[\lambda m_{h}^{2}+\frac{e^{2}}{\left(4\pi\right)^{2}}3p^{2}+\frac{\lambda}{2}m_{h}^{2}+3e^{2}m^{2}\right]\left(\frac{2}{\varepsilon}-\gamma+\ln\left(4\pi\right)\right) \;, \label{rhorhodiv} 
\end{eqnarray}
we obtain

\begin{eqnarray}
Z_{h}^{\left(1\right)} & = & \frac{3e^{2}}{16\pi^{2}}\left(\frac{2}{\varepsilon}-\gamma+\ln\left(4\pi\right)\right) \;, \label{zrho}
\end{eqnarray}
and 

\begin{equation}
 (\delta \sigma)^{\left(1\right)}_{\rm div}  =  \frac{1}{\left(4\pi\right)^{2}}\frac{1}{v^{2}}\left[3e^{2}m^{2}+\frac{3\lambda}{2}m_{h}^{2}\right]\left(\frac{2}{\varepsilon}-\gamma+\ln\left(4\pi\right)\right) \;, \label{dsrho} 
\end{equation}
agreeing with the previous result, eq.(\ref{divtad}). Therefore, the renormalized two point function $\left\langle \rho \rho\right\rangle^{\rm 1-loop}$
turns out to be 

\begin{eqnarray}
\left\langle \rho \rho\right\rangle^{\rm 1-loop}(p^2)   & = & \frac{1}{p^{2}}+
\frac{1}{\left(p^{2}\right)^{2}}\Pi_{\rho\rho}^{(1)}\left(p^{2}\right) \;, \label{rrhorho}
\end{eqnarray}
where

\begin{eqnarray}
\Pi_{\rho\rho}^{(1)} & = & -\frac{\lambda}{\left(4\pi\right)^{2}}m_{h}^{2}\left[-\left(\ln\left(\frac{m_{h}^{2}}{\mu^{2}}\right)-1\right)+\int_{0}^{1}dx\ln\left(\frac{H\left(0,m_{h}^{2},p^{2}\right)}{\mu^{2}}\right)\right] \nonumber \\
 &  & +4\frac{e^{2}}{\left(4\pi\right)^{2}}p^{2}\int_{0}^{1}dx\left\{ -\ln\left(\frac{H\left(m^{2},m_{h}^{2},p^{2}\right)}{\mu^{2}}\right)\right. \nonumber \\
 &  & +\frac{p^{2}}{m^{2}}\left(1-x\right)^{2}\left[\ln\left(\frac{H\left(0,m_{h}^{2},p^{2}\right)}{\mu^{2}}\right)-\ln\left(\frac{H\left(m^{2},m_{h}^{2},p^{2}\right)}{\mu^{2}}\right)\right] \nonumber \\
 &  & \left.-\frac{1}{2m^{2}}\left[H\left(0,m_{h}^{2},p^{2}\right)\left(\ln\left(\frac{H\left(0,m_{h}^{2},p^{2}\right)}{\mu^{2}}\right)-1\right)-H\left(m^{2},m_{h}^{2},p^{2}\right)\left(\ln\left(\frac{H\left(m^{2},m_{h}^{2},p^{2}\right)}{\mu^{2}}\right)-1\right)\right]\right\} \nonumber \\[3mm] 
 \label{pirhorho}
\end{eqnarray}

For the resummed one-loop two-point correlation function we get

\begin{eqnarray}
\left\langle \rho\left(p\right)\rho\left(-p\right)\right\rangle^{\rm 1-loop}(p^2)  & = & \frac{1}{p^{2}-\Pi_{\rho\rho}^{(1)}\left(p^{2}\right)} + O(\hbar^2) \;. \label{rhorhors}
\end{eqnarray}
Since
\begin{equation} 
H\left(0,m_{h}^{2},0\right)=\left(1-x\right)m_{h}^{2} \;, \label{Hr}
\end{equation} 
it follows that 
\begin{eqnarray}
\int_{0}^{1}dx\ln\left(\frac{H\left(0,m_{h}^{2},0\right)}{\mu^{2}}\right) & = & \ln\left(\frac{m_{h}^{2}}{\mu^{2}}\right)-1 \;, \label{Hr1}
\end{eqnarray}
implying that
\begin{equation} 
 \Pi_{\rho\rho}^{(1)} \left(0\right)=0 \;. \label{mrhorho}
\end{equation} 
This shows that, as required by the Ward identity (\ref{gboWq}), the Goldstone field $\rho$ remains massless.  

\section{Appendix C: Green's functions of the  composite operators \label{C}}
Let us now face the computation of the correlation functions with insertions of the local composite operators $(O(x),V_\mu(x))$.

\subsection{Evaluation of the one-loop connected Green function $\left\langle h\left(x\right)O\left(y\right)\right\rangle $}

\begin{figure}[h]
\begin{center}
\includegraphics[width=.8\textwidth,angle=0]{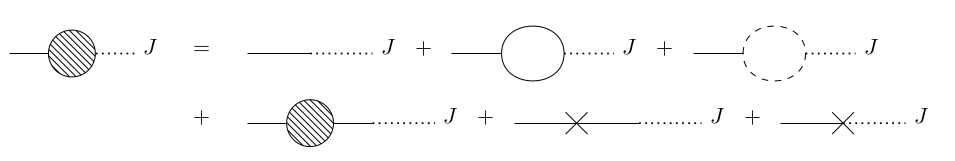}
\caption{
Feynman diagrams that contribute at one-loop to the two-point Green's function $\left\langle h\left(x\right)\right\rangle _{J}$.
}
\label{hJ}
\end{center}
\end{figure}

As explained in Section 5, in order to evaluate $\left\langle h\left(x\right)O\left(y\right)\right\rangle $, we first look at  $\left\langle h\left(x\right)\right\rangle _{J}$, where the source $J$ is an external field. At the end we shall differentiate with respect to $J$ and set it to $0$. The Feynman diagrams which contribute to $\left\langle h\left(x\right)\right\rangle _{J}$ are given in Figure \ref{hJ}. 
% \begin{eqnarray}
% \feynmandiagram [layered layout, horizontal=a to b] {a -- b [blob],b -- [ghost]c[particle=\(J\)]
% };  & = & \feynmandiagram [layered layout, horizontal=a to b] {
% a --  b
% -- [ half left] c, c --[half left] b --[ghost], c --[ghost] d [particle=\(J\)],};+\feynmandiagram [layered layout, horizontal=a to b] {
% a --  b
% -- [ scalar, half left] c, c --[scalar, half left] b,c -- [ghost] d [particle=\(J\)]}; \nonumber \\
% & & + \feynmandiagram [layered layout, horizontal=a to b] {
% a -- b -- [ghost] c [particle=\(J\)],};+\feynmandiagram [layered layout, horizontal=a to b] {a -- b [blob],b -- c,c -- [ghost]d[particle=\(J\)]
% };\nonumber \\
% & & + \feynmandiagram [layered layout, horizontal=a to b] {
% a --  [insertion=1] b,b -- c,c --[ghost] d [particle=\(J\)],};+ \feynmandiagram [layered layout, horizontal=a to b] {
% a --[insertion=1] b -- [ghost] c [particle=\(J\)],};
% \end{eqnarray}
For each contribution we get 
\begin{eqnarray}
\left\langle h\left(p\right)\right\rangle _{J}^{1} & = & \frac{1}{2}\frac{1}{p^{2}+m_{h}^{2}}\left(-3\lambda v\right)\left(-\widetilde{J}\left(p\right)\right)\int\frac{d^{d}k}{\left(2\pi\right)^{d}}\frac{1}{k^{2}+m_{h}^{2}}\frac{1}{\left(p-k\right)^{2}+m_{h}^{2}} \;, \label{hj1}
\end{eqnarray}

\begin{eqnarray}
\left\langle h\left(p\right)\right\rangle _{J}^{2} & = & \frac{1}{2}\frac{1}{p^{2}+m_{h}^{2}}\left(-\lambda v\right)\left(-\widetilde{J}\left(p\right)\right)\int\frac{d^{d}k}{\left(2\pi\right)^{d}}\frac{1}{k^{2}}\frac{1}{\left(p-k\right)^{2}} \;, \label{hj2} 
\end{eqnarray}

\begin{eqnarray}
\left\langle h\left(p\right)\right\rangle _{J}^{3} & = & \frac{1}{p^{2}+m_{h}^{2}}\left(-v\widetilde{J}\left(p\right)\right) \;, \label{hj3}
\end{eqnarray}

\begin{eqnarray}
\left\langle h\left(p\right)\right\rangle _{J}^{4} & = & \frac{1}{\left(p^{2}+m_{h}^{2}\right)^{2}}\Pi_{hh}^{\rm 1-loop}\left(p^{2}\right)\left(-v\widetilde{J}\left(p\right)\right) \;, \label{hj4}
\end{eqnarray}
where $\Pi_{hh}^{\rm 1-loop}$ stands for the Higgs one-loop self energy, see previous Appendix \ref{B} and \cite{Dudal:2019aew,Dudal:2019pyg} 
\begin{eqnarray}
\left\langle h\left(p\right)\right\rangle _{J}^{5} & = & \frac{1}{\left(p^{2}+m_{h}^{2}\right)^{2}}\left(-\left(Z_{h}^{\left(1\right)}p^{2}+\left(Z_{\lambda}^{\left(1\right)}+Z_{h}^{\left(1\right)}\right)m_{h}^{2}+(\delta \sigma)^{\left(1\right)}v^{2}\right)\right)\left(-v\widetilde{J}\left(p\right)\right) \;, \label{hj5}
\end{eqnarray}

\begin{eqnarray}
\left\langle h\left(p\right)\right\rangle _{J}^{6} & = & \frac{1}{p^{2}+m_{h}^{2}}\left(-v\left(Z_{J}^{\left(1\right)}+Z_{h}^{\left(1\right)}\right)\widetilde{J}\left(p\right)\right) \;. \label{hj6}
\end{eqnarray}
Therefore, collecting all contributions, differentiating with respect to the source $J$ and setting it to $0$, for the one-loop correlation function $\left\langle h\left(p\right)O\left(-p\right)\right\rangle^{\rm 1-loop}$, we get 

\begin{eqnarray}
\left\langle h\left(p\right)O\left(-p\right)\right\rangle^{\rm 1-loop} & = & \frac{1}{p^{2}+m_{h}^{2}}\left(\frac{3}{2}\lambda v\right)\int\frac{d^{d}k}{\left(2\pi\right)^{d}}\frac{1}{k^{2}+m_{h}^{2}}\frac{1}{\left(p-k\right)^{2}+m_{h}^{2}} \nonumber  \\
 &  & +\frac{1}{p^{2}+m_{h}^{2}}\left(\frac{1}{2}\lambda v\right)\int\frac{d^{d}k}{\left(2\pi\right)^{d}}\frac{1}{k^{2}}\frac{1}{\left(p-k\right)^{2}}\nonumber \\
 &  & +\left(-v\right)\frac{1}{p^{2}+m_{h}^{2}}\left(Z_{JJ}^{\left(1\right)}+Z_{h}^{\left(1\right)}\right)\nonumber \\
 &  & + \frac{-v}{p^{2}+m_{h}^{2}}+\frac{-v}{\left(p^{2}+m_{h}^{2}\right)^{2}}\left(\Pi_{hh}^{1-loop}\left(p^{2}\right)-\left(Z_{h}^{\left(1\right)}p^{2}+\left(Z_{\lambda}^{\left(1\right)}+Z_{h}^{\left(1\right)}\right)m_{h}^{2}+(\delta \sigma)^{\left(1\right)}v^{2}\right)\right) \,. \nonumber \\
  \label{ho1}
 %& = & \widetilde{J}\left(p\right)\frac{1}{p^{2}+m_{h}^{2}}v\left[\frac{3}{2}\lambda\int\frac{d^{d}k}{\left(2\pi\right)^{d}}\frac{1}{k^{2}+m_{h}^{2}}\frac{1}{\left(p-k\right)^{2}+m_{h}^{2}}+\frac{1}{2}\lambda\int\frac{d^{d}k}{\left(2\pi\right)^{d}}\frac{1}{k^{2}}\frac{1}{\left(p-k\right)^{2}}-\left(Z_{J}^{\left(1\right)}+Z_{h}^{\left(1\right)}\right)\right]\\
% &  & +\left(-v\widetilde{J}\left(p\right)\right)\left\langle h\left(p\right)h\left(-p\right)\right\rangle _{R}
\end{eqnarray}
After isolating the divergent part, we have 

\begin{eqnarray*}
Z_{JJ}^{\left(1\right)}+Z_{h}^{\left(1\right)} & = & 2\frac{\lambda}{\left(4\pi\right)^{2}}\left(\frac{2}{\varepsilon}-\gamma+\ln\left(4\pi\right)\right).
\end{eqnarray*}
Using the previous result for $Z_{h}^{\left(1\right)}$, eq.(\ref{zhhf}), it turns out that  
\begin{eqnarray}
Z_{JJ}^{\left(1\right)} & = & \frac{1}{16\pi^{2}}\left(2\lambda-3e^{2}\right)\left(\frac{2}{\varepsilon}-\gamma+\ln\left(4\pi\right)\right) 
\;. \label{ZJJ1}
\end{eqnarray}
As was already mentioned, this result is in perfect agreement with the Ward identities,
\begin{equation}
Z_{JJ}^{\left(1\right)}=Z_{\lambda}^{\left(1\right)}+Z_{h}^{\left(1\right)}-2\frac{\left(\delta\sigma\right)^{\left(1\right)}_{\rm div}}{\lambda}\,.
\end{equation}

\subsection{Evaluation of $\left\langle A_{\mu}\left(p\right)V_{\nu}\left(-p\right)\right\rangle $}
\begin{figure}[h]
\begin{center}
\includegraphics[width=.8\textwidth,angle=0]{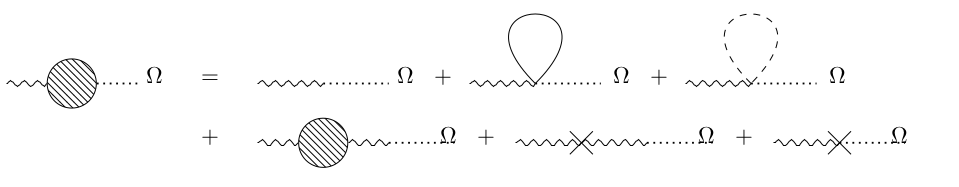}
\caption{One-loop diagrams contributing to the Green function $\left\langle A_{\mu}\left(p\right)\right\rangle _{\Omega}$  .
}
\label{phomega}
\end{center}
\end{figure}

Let us consider finally the last Green function, $\left\langle A_{\mu}\left(p\right)V_{\nu}\left(-p\right)\right\rangle $, whose contributing Feynman diagrams at one-loop order are given in the Figure \ref{phomega}. 
% \begin{eqnarray*}
% \feynmandiagram [layered layout, horizontal=a to b] {a -- [photon] b [blob],b -- [ghost]c[particle=\(\Omega\)]
% };  & = & \feynmandiagram [layered layout, horizontal=a to b] {
% a -- [photon]  b
% -- [photon, half left] c, c --[half left] b --[ghost], c --[ghost] d [particle=\(\Omega\)],};+\feynmandiagram [layered layout, horizontal=a to b] {
% a --[photon]  b
% -- [ half left] c, c --[scalar, half left] b,c -- [ghost] d [particle=\(\Omega\)]};  \\
% & & + \feynmandiagram [layered layout, horizontal=a to b] {
% a --[photon] b-- [out=135, in=45, loop, min distance=2cm] b -- [ghost] c [particle=\(\Omega\)],}; + \feynmandiagram [layered layout, horizontal=a to b] {
% a --[photon] b-- [scalar, out=135, in=45, loop, min distance=2cm] b -- [ghost] c [particle=\(\Omega\)],};+\feynmandiagram [layered layout, horizontal=a to b] {a --[photon] b [blob],b --[photon] c,c -- [ghost]d[particle=\(\Omega\)]
% }; \\
% & & + \feynmandiagram [layered layout, horizontal=a to b] {
% a --  [photon, insertion=1] b,b --[photon] c,c --[ghost] d [particle=\(\Omega\)],};+\feynmandiagram [layered layout, horizontal=a to b] {
% a --[photon] b -- [ghost] c [particle=\(\Omega\)],};+\feynmandiagram [layered layout, horizontal=a to b] {
% a --[photon,insertion=1] b -- [ghost] c [particle=\(\Omega\)],};   
% \end{eqnarray*}

Each contribution turns out to be 

\begin{eqnarray}
\left\langle A_{\mu}\left(p\right)\right\rangle _{\Omega}^{1} & = & \frac{P_{\mu\alpha}\left(p\right)}{p^{2}+m^{2}}\left(-2e^{2}v\delta_{\alpha\beta}\right)\left(-ev\widetilde{\Omega}_{\gamma}\left(p\right)\right)\int\frac{d^{d}k}{\left(2\pi\right)}\frac{P_{\beta\gamma}\left(k\right)}{k^{2}+m^{2}}\frac{1}{k^{2}+m_{h}^{2}} \nonumber \\
 & = & \frac{P_{\mu\alpha}}{p^{2}+m^{2}}\left(2e^{3}v^{2}\right)\widetilde{\Omega}_{\alpha}\left(p\right) \nonumber \\
 &  & \frac{1}{\left(4\pi\right)^{\frac{d}{2}}}\int_{0}^{1}dx\left\{ \Gamma\left(2-\frac{d}{2}\right)H\left(m^{2},m_{h}^{2}\right)^{\frac{d}{2}-2}\right. \nonumber \\
 &  & \left.-\frac{1}{2m^{2}}\Gamma\left(1-\frac{d}{2}\right)\left(H\left(0,m_{h}^{2}\right)^{\frac{d}{2}-1}-H\left(m^{2},m_{h}^{2}\right)^{\frac{d}{2}-1}\right)\right\} \;, \label{av1}
\end{eqnarray}

\begin{eqnarray}
\left\langle A_{\mu}\left(p\right)\right\rangle _{\Omega}^{2} & = & \frac{P_{\mu\alpha}\left(p\right)}{p^{2}+m^{2}}\int\frac{d^{d}k}{\left(2\pi\right)^{d}}ie\left(-2k_{\alpha}+p_{\alpha}\right)\frac{1}{k^{2}+m_{h}^{2}}\left(-\frac{1}{2}i\left(p_{\beta}-2k_{\beta}\right)\widetilde{\Omega}_{\beta}\left(p\right)\right)\frac{1}{\left(p-k\right)^{2}} \nonumber \\
 & = & \frac{P_{\mu\alpha}\left(p\right)}{p^{2}+m^{2}}\left(e\widetilde{\Omega}_{\alpha}\left(p\right)\right)\left\{ \Gamma\left(1-\frac{d}{2}\right)\frac{1}{\left(4\pi\right)^{\frac{d}{2}}}\int_{0}^{1}dxH\left(m_{h}^{2},0\right)^{\frac{d}{2}-1}\right\} \;, \label{av2}
\end{eqnarray}

\begin{eqnarray}
\left\langle A_{\mu}\left(p\right)\right\rangle _{\Omega}^{3} & = & \frac{1}{2}\frac{P_{\mu\alpha}\left(p\right)}{p^{2}+m^{2}}\left(-e\widetilde{\Omega}_{\alpha}\left(p\right)\right)\int\frac{d^{d}k}{\left(2\pi\right)^{d}}\frac{1}{k^{2}+m_{h}^{2}} \nonumber \\
 & = & \frac{P_{\mu\alpha}\left(p\right)}{p^{2}+m^{2}}\left(-\frac{1}{2}e\widetilde{\Omega}_{\alpha}\left(p\right)\right)\frac{1}{\left(4\pi\right)^{\frac{d}{2}}}\Gamma\left(1-\frac{d}{2}\right)\left(m_{h}^{2}\right)^{\frac{d}{2}-1} \;, \label{av3} 
\end{eqnarray}

\begin{eqnarray}
\left\langle A_{\mu}\left(p\right)\right\rangle _{\Omega}^{4}  =  \frac{1}{2}\frac{P_{\mu\alpha}\left(p\right)}{p^{2}+m^{2}}\left(-e\widetilde{\Omega}_{\alpha}\left(p\right)\right)\int\frac{d^{d}k}{\left(2\pi\right)^{d}}\frac{1}{k^{2}}
  =  0 \;, \label{av4}
\end{eqnarray}

\begin{eqnarray}
\left\langle A_{\mu}\left(p\right)\right\rangle _{\Omega}^{5} & = & \frac{P_{\mu\alpha}\left(p\right)}{\left(p^{2}+m^{2}\right)^{2}}\Pi^{1-loop}_{AA}\left(p\right)\left(-\frac{1}{2}ev^{2}\widetilde{\Omega}_{\alpha}\left(p\right)\right) \;, \label{av5}
\end{eqnarray}
where $\Pi^{1-loop}_{AA}$ stands for the one-loop gauge boson self-energy, see eq.(\ref{aa2p}),  
\begin{eqnarray}
\left\langle A_{\mu}\left(p\right)\right\rangle _{\Omega}^{6} & = & \frac{P_{\mu\alpha}\left(p\right)}{\left(p^{2}+m^{2}\right)^{2}}\left(-\left(Z_{A}^{\left(1\right)}\left(p^{2}\delta_{\alpha\beta}-p_{\alpha}p_{\beta}\right)+Z_{h}^{\left(1\right)}m^{2}\delta_{\alpha\beta}\right)\right)\left(-\frac{1}{2}ev^{2}\widetilde{\Omega}_{\beta}\left(p\right)\right) \nonumber \\
 & = & \frac{P_{\mu\alpha}\left(p\right)}{\left(p^{2}+m^{2}\right)^{2}}\left(-\left(p^{2}Z_{A}^{\left(1\right)}+Z_{h}^{\left(1\right)}m^{2}\right)\right)\left(-\frac{1}{2}ev^{2}\widetilde{\Omega}_{\alpha}\left(p\right)\right) \;, \label{av6}
\end{eqnarray}

\begin{eqnarray}
\left\langle A_{\mu}\left(p\right)\right\rangle _{\Omega}^{7} & = & \frac{P_{\mu\alpha}\left(p\right)}{p^{2}+m^{2}}\left(-\frac{1}{2}ev^{2}\widetilde{\Omega}_{\alpha}\left(p\right)\right) \;, \label{av7} 
\end{eqnarray}
\begin{eqnarray}
\left\langle A_{\mu}\left(p\right)\right\rangle _{\Omega}^{8} & = & \frac{P_{\mu\alpha}\left(p\right)}{p^{2}+m^{2}}\left[-\frac{1}{2}\left(Z_{\Omega\Omega}^{\left(1\right)}+Z_{h}^{\left(1\right)}\right)ev^{2}\widetilde{\Omega}_{\alpha}\left(p\right)+Z_{\Upsilon\Omega}^{\left(1\right)}p^{2}P_{\alpha\nu}\left(p\right)\widetilde{\Omega}_{\nu}\left(p\right)\right] \;. \label{av8}
\end{eqnarray}
Therefore, after summing up all contributions, differentiating  with respect to $\Omega_\nu$ and setting it to zero, for the divergent part of $\left\langle A_{\mu}\left(p\right) V_\nu(-p) \right\rangle _{\rm div}^{\rm 1-loop}$, we obtain 

%\begin{eqnarray*}
%\left\langle A_{\mu}\left(p\right)\right\rangle _{\Omega} & = & \frac{P_{\mu\alpha}}{p^{2}+m^{2}}\left(2e^{3}v^{2}\right)\widetilde{\Omega}_{\alpha}\left(p\right)\\
% &  & \frac{1}{\left(4\pi\right)^{\frac{d}{2}}}\int_{0}^{1}dx\left\{ \Gamma\left(2-\frac{d}{2}\right)H\left(m^{2},m_{h}^{2}\right)^{\frac{d}{2}-2}\right.\\
% &  & \left.-\frac{1}{2m^{2}}\Gamma\left(1-\frac{d}{2}\right)\left(H\left(0,m_{h}^{2}\right)^{\frac{d}{2}-1}-H\left(m^{2},m_{h}^{2}\right)^{\frac{d}{2}-1}\right)\right\} \\
% &  & +\frac{P_{\mu\alpha}\left(p\right)}{p^{2}+m^{2}}\left(e\widetilde{\Omega}_{\alpha}\left(p\right)\right)\left\{ \Gamma\left(1-\frac{d}{2}\right)\frac{1}{\left(4\pi\right)^{\frac{d}{2}}}\int_{0}^{1}dxH\left(m_{h}^{2},0\right)^{\frac{d}{2}-1}\right\} \\
% &  & +\frac{P_{\mu\alpha}\left(p\right)}{p^{2}+m^{2}}\left(-\frac{1}{2}e\widetilde{\Omega}_{\alpha}\left(p\right)\right)\frac{1}{\left(4\pi\right)^{\frac{d}{2}}}\Gamma\left(1-\frac{d}{2}\right)\left(m_{h}^{2}\right)^{\frac{d}{2}-1}\\
% &  & +\frac{P_{\mu\alpha}\left(p\right)}{p^{2}+m^{2}}\left[-\frac{1}{2}ev^{2}\widetilde{\Omega}_{\alpha}\left(p\right)\left(Z_{\Omega\Omega}^{\left(1\right)}+Z_{h}^{\left(1\right)}\right)+Z_{\Upsilon\Omega}^{\left(1\right)}p^{2}\widetilde{\Omega}_{\alpha}\left(p\right)\right]\\
% &  & +\left(-\frac{1}{2}ev^{2}\widetilde{\Omega}_{\alpha}\left(p\right)\right)\left\langle A_{\mu}\left(p\right)A_{\alpha}\left(-p\right)\right\rangle _{R}
%\end{eqnarray*}

\begin{eqnarray}
\left\langle A_{\mu}\left(p\right) V_\nu(-p) \right\rangle _{\rm div}^{\rm 1-loop} & = & \frac{P_{\mu\nu}}{p^{2}+m^{2}}\left(\frac{2}{\varepsilon}-\gamma+\ln\left(4\pi\right)\right)\frac{1}{\left(4\pi\right)^{2}}\left\{ \frac{3}{2}em^{2}-e\frac{p^{2}}{6}\right\} \nonumber \\
 &  & +\frac{P_{\mu\nu}}{p^{2}+m^{2}}\left\{ -\frac{1}{2}ev^{2}\left(Z_{\Omega\Omega}^{\left(1\right)}+Z_{h}^{\left(1\right)}\right)+Z_{\Upsilon\Omega}^{\left(1\right)}p^{2}\right\} \;. \label{divav}
\end{eqnarray}
It turns out thus that 
\begin{equation} 
Z_{\Omega\Omega}^{\left(1\right)}+Z_{h}^{\left(1\right)}= \frac{3e^2}{16\pi^2}\left(\frac{2}{\varepsilon}-\gamma+\ln\left(4\pi\right)\right) \;, \label{1ck}
\end{equation} 
and
\begin{equation}
 Z_{\Upsilon\Omega}^{\left(1\right)}  =  \frac{e}{96\pi^{2}}\left(\frac{2}{\varepsilon}-\gamma+\ln\left(4\pi\right)\right)  = - \frac{1}{2e} Z_{AA}^{(1)}  \;. \label{2ck}
\end{equation}
One sees that the factor $Z_{h}^{\left(1\right)}$, eq.(\ref{zhhf}), cancels completely the right hand side of eq.(\ref{1ck}), yielding 
\begin{equation} 
 Z_{\Omega\Omega}^{\left(1\right)}= 0 \;. \label{3ck}
\end{equation} 
Equations (\ref{2ck}) and (\ref{3ck}) are in full agreement with the Ward identities.
%\begin{eqnarray*}
%Z_{\Upsilon\Omega}^{\left(1\right)} & = & \frac{e}{96\pi^{2}}\left(\frac{2}{\varepsilon}-\gamma+\ln\left(4\pi\right)\right)\\
%Z_{\Omega\Omega}^{\left(1\right)} & = & 0.
%\end{eqnarray*}

\end{document}